\documentclass[modern]{aastex63}
\newcommand{\kms}{\ensuremath{\,\rm{km\,s^{-1}}}}
\newcommand{\cmk}{\ensuremath{\,\rm{cm^{-3}K}}}
\newcommand{\lt}{\ensuremath{<}} 
\newcommand{\gt}{\ensuremath{>}} 
\shorttitle{ISM Thermal Pressures away from Stars}
\shortauthors{Jenkins \& Tripp}
\usepackage[utf8]{inputenc}
\usepackage{epsfig}
\begin{document}
\title{Thermal Pressures in the Interstellar Medium away from Stellar Environments 
\footnote{Based on observations with the NASA/ESA Hubble Space Telescope obtained from 
the Data Archive at the Space Telescope Science Institute, which is operated by the Associations 
of Universities for Research in Astronomy, Incorporated, under NASA contract NAS5-26555.~ 
\copyright 2021. The American Astronomical Society. All rights reserved.}}
\author[0000-0003-1892-4423]{Edward B. Jenkins}
\affiliation{Department of Astrophysical Sciences, Princeton University,
Princeton, NJ 08544-1001}
\author[0000-0002-1218-640X]{Todd M. Tripp}
\affiliation{Dept. of Astronomy, University of Massachusetts, 710 North Pleasant Street, 
Amherst, MA 01003-9305, USA}
\correspondingauthor{E. B. Jenkins}
\email{ebj@astro.princeton.edu}
\email{ttripp@umass.edu}
\begin{abstract}

Interstellar thermal pressures can be measured using \ion{C}{1} absorption lines that probe the 
pressure-sensitive populations of the fine-structure levels of its ground state. In a survey of 
\ion{C}{1}  absorption toward Galactic hot stars, Jenkins \& Tripp (2011) found evidence of 
small amounts ($\sim 0.05\%$) of gas at high pressures ($p/k \gg 10^4\cmk$) mixed with a 
more general presence of lower pressure material exhibiting a log normal distribution that 
spanned the range $10^3 \lesssim p/k \lesssim 10^4\cmk$. In this paper, we study Milky Way 
\ion{C}{1}  lines in the spectra of extragalactic sources instead of Galactic stars and thus 
measure the pressures without being influenced by regions where stellar mass loss and 
\ion{H}{2} region expansions could create localized pressure elevations.  We find that the 
distribution of low pressures in the current sample favors slightly higher pressures than the 
earlier survey, and the fraction of gaseous material at extremely high pressures is about the 
same as that found earlier.  Thus we conclude that the earlier survey was not appreciably 
influenced by the stellar environments, and the small amounts of high pressure gas indeed exist 
within the general interstellar medium.
\end{abstract}
\section{Introduction}\label{sec:intro}
In the local part of our Galaxy, pressures within the interstellar medium (ISM) are manifested in 
several mutually interacting forms: thermal ($nkT$), magnetic ($B^2/8\pi$), dynamical or 
turbulent ($\rho v^2$), and from cosmic rays.  The overall average of the combined pressure of 
gaseous material that is not self gravitating, amounting to a Galactic midplane pressure 
$p\approx 3.9\times 10^{-12} {\rm dyne~cm^{-2}}$ (or $p/k\approx 3\times 10^4\cmk$), is 
balanced against the weight of material in the plane’s gravitational potential (Boulares \& Cox 
1990 ; Lockman \& Gehman 1991 ; Koyama \& Ostriker 2009).  In turn, these pressures directly 
influence the scale height of the ISM on either side of the plane (McKee 1990).  Thermal 
pressures are a minor portion of the total pressure ($\sim 15\%$), but as we discuss below, 
thermal pressure surveys have revealed some surprising results that provide insights on the 
structure and physics of the ISM.  In the absence of gravitational binding or ephemeral positive 
and negative excursions caused by dynamical processes, we expect to find that an acceptable 
range for these pressures is regulated by the heating and cooling rates for the neutral ISM that 
define a thermal instability that creates two separate, coexisting phases\footnote{We ignore 
here two other phases, the warm ionized medium (Haffner et al. 2009 ; Geyer \& Walker 2018) 
and a very hot phase at temperatures $10^6 \lt T \lt 10^7$\,K (Spitzer 1990).}, the warm 
neutral medium (WNM; $T\sim 10^4$\,K) and a cold neutral medium (CNM; $T\sim 100$\,K) 
(Field 1965 ; Field et al. 1969 ; Wolfire et al. 2003).

The distribution of thermal pressures within and slightly outside the range permitted by the 
thermal instability, $2-5\times 10^3\cmk$, offers insights on the strengths of deviations caused 
by dynamical processes, such as shocks or turbulence.  Simulations of these turbulent regimes 
reveal that thermal pressures should generally conform to a log-normal distribution, with a 
width proportional to the sonic  Mach number and a mild skewness that depends on the 
effective equation of state of the gas and the mix between solenoidal (divergence-free) and 
compressive (curl-free) forcing (Federrath 2013 ; Kim et al. 2013 ; Gazol 2014 ; Kritsuk et al. 
2017 ; Mocz \& Burkhart 2019).   

Observationally, we can measure thermal pressures in the CNM by comparing absorption 
features of \ion{C}{1} in three levels of fine-structure excitation, $^3{\rm P}_0$, $^3{\rm P}_1$, 
and $^3{\rm P}_3$ of the ground electronic state ${\rm (1s^2)2s^22p^2}$, as viewed by their 
multiple absorption features in the ultraviolet spectra of background stars (Jenkins \& Tripp 
2001).  Hereafter, we will refer to column densities of carbon atoms in these three states as 
$N$(\ion{C}{1}) (lowest 
level with $J=0$), $N$(\ion{C}{1}$^*$) ($J=1$ level with an excitation energy $E/k=23.6$\,K) 
and $N$(\ion{C}{1}$^{**}$) ($J=2$ level with an excitation energy $E/k=62.4$\,K).  The total 
amount of \ion{C}{1} in all three levels, 
$N$(\ion{C}{1})$+N$(\ion{C}{1}$^*)+N$(\ion{C}{1}$^{**})$, will be stated as 
$N$(\ion{C}{1})$_{\rm total}$.  The three levels are both populated and depopulated by 
collisions with other particles, such as atoms, electrons, and molecules, along with optical 
pumping by starlight (Silva \& Viegas 2002).  The relative fractions that are found for these 
levels indicate local densities and temperatures, since the collisional interactions compete with 
spontaneous radiative decays.

Measurements of \ion{C}{1} absorptions from the three levels toward bright stars in the local 
part of our Galaxy have a long history, starting with a survey of \ion{C}{1} features observed 
during the first year of operation of the Copernicus satellite (Jenkins \& Shaya 1979), followed 
by a later, more extensive study of Copernicus data by Jenkins et al. (1983).  The ability to 
observe fainter stars using the International Ultraviolet Explorer (IUE) and the Hubble Space 
Telescope (HST) brought about investigations of sight lines through supernova remnants, where 
strongly elevated pressures caused by shock waves in the ISM could be sensed  (Jenkins et al. 
1981 ; Jenkins et al. 1984 ; Jenkins \& Wallerstein 1995 ; Jenkins et al. 1998 ; Ritchey et al. 
2020).   Meyer et al. (2012) detected extraordinarily high pressures in a nearby, cold cloud at a 
distance $\sim 20$~pc from the Sun (Peek et al. 2011).  Welty et al. (2016) and Roman-Duval et 
al. (2021) have used \ion{C}{1} to measure thermal pressures for gas in the Magellanic Clouds, 
and there have been a number of investigations of \ion{C}{1} excitations in distant damped Ly-$\alpha$
(DLA) systems (Quast et al. 2002 ; Srianand et al. 2005 ; Jorgenson et al. 2010 ; 
Noterdaeme et al. 2010 ; Carswell et al. 2011 ; Ma et al. 2015 ; Noterdaeme et al. 2015 ; 
Balashev et al. 2020 ; Klimenko \& Balashev 2020).  For the most distant of these systems one 
must include the effect of the cosmic microwave background (CMB), which exposes the carbon 
atoms to black body radiation at a temperature $2.725(1+z_{\rm abs})$ from all directions in 
the sky.

We now focus on \ion{C}{1} pressure measurements toward 89 stars in the local region of our 
Galaxy reported by Jenkins \& Tripp (2011, hereafter JT11), who obtained spectra from the 
Mikulsky Archive for Space Telescopes (MAST) recorded by the highest resolution echelle mode 
of the Space Telescope Imaging Spectrograph (STIS) on HST.  This program supplemented an 
earlier STIS survey of \ion{C}{1} for a limited sample of targeted sightlines by Jenkins \& Tripp 
(2001).  To determine the three separate column densities as a function of velocity, for each 
spectrum they solved a large set of simultaneous linear equations based on the absorption 
profiles of many different multiplets, each of which usually had overlapping features.  This 
analysis method is described in the earlier paper by Jenkins \& Tripp (2001). 

An efficient way to interpret results from not only single regions at a specific pressure but also 
the effect of a superposition of regions at different pressures at a common velocity is through a 
diagram that plots the outcomes for $f1=N$(\ion{C}{1}$^*$)/$N$(\ion{C}{1})$_{\rm total}$ on 
the $x$-axis and $f2=N$(\ion{C}{1}$^{**}$)/$N$(\ion{C}{1})$_{\rm total}$ on the $y$-axis.  A 
result for $f1$ and $f2$ that arises from a single region at a given pressure should fall 
somewhere on a curve computed for the respective excitations at various pressures and 
temperatures.  Outcomes from two or more regions with different pressures will reside on a 
\ion{C}{1}-weighted ``center of mass’’ location that can be displaced from these curves, as 
illustrated in Fig.~5 of Jenkins \& Tripp (2001).  This concept was introduced in the early paper 
by Jenkins \& Shaya (1979) and repeated in many of the subsequent papers on \ion{C}{1} 
observations referenced above.  A more detailed discussion of the implications of the displaced 
points was covered in Section~5 of JT11, which presented evidence that most lines of sight 
were dominated by \ion{C}{1} at pressures in the range $3\lesssim \log(p/k)_{\rm low} \lesssim 
4$ and with fractional contributions to the total $g_{\rm low}\approx 0.95$ for most but not all 
cases.  The remaining usually small contributions $1-g_{\rm low}$ came from gas at 
extraordinarily high pressures ($\log (p/k)_{\rm high}\gt 5.5$).  After correcting for the shifts in 
the ionization equilibrium that favored the neutral form of carbon at high pressures, JT11 
calculated that this high pressure contribution represented a relative mass fraction equal to 
$5\times 10^{-4}$.  
\newpage
\section{Possible Origins for the High Pressure Gas}\label{sec:origin}

\begin{figure}
\plotone{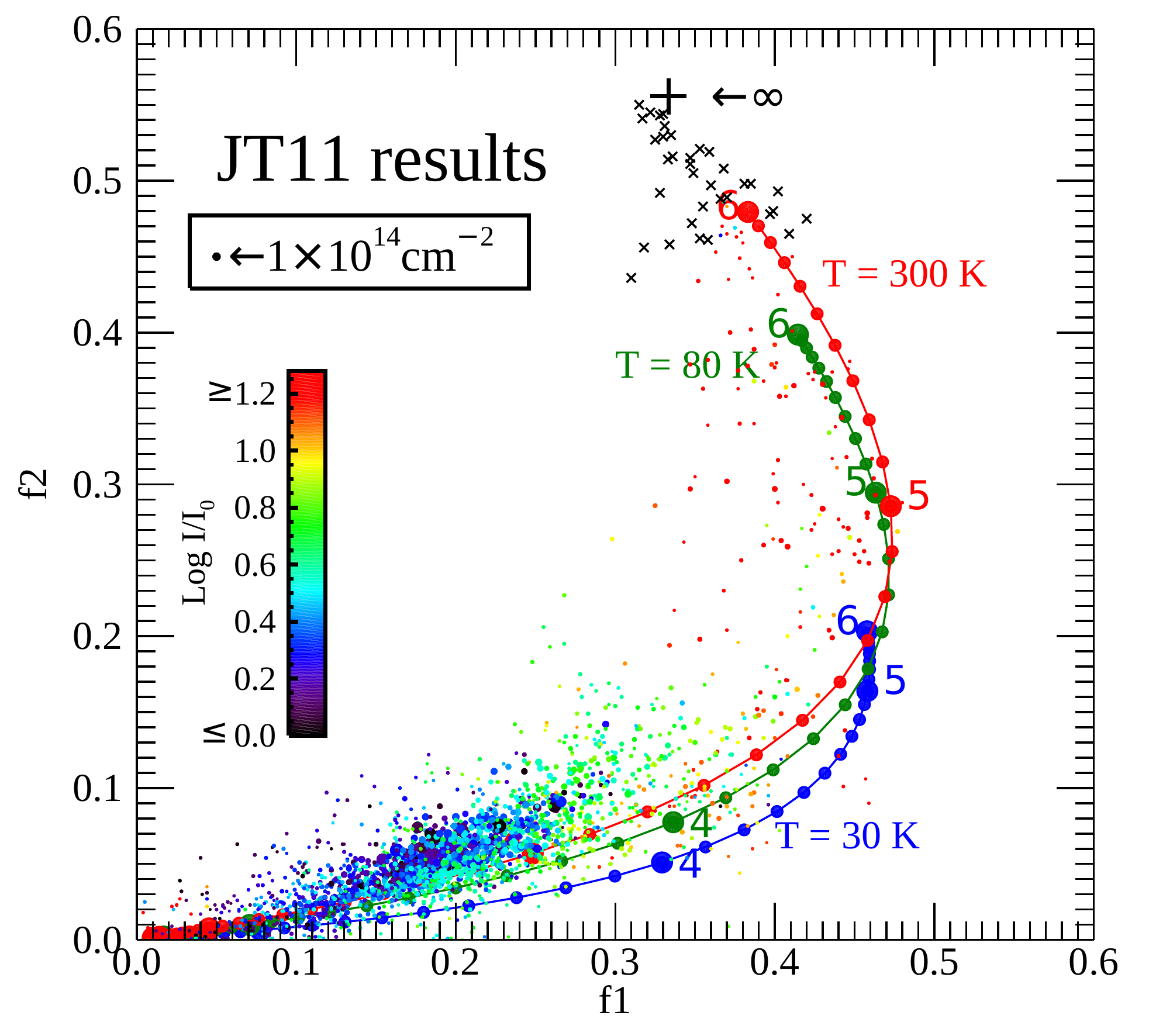}
\caption{A depiction of pressure indicators in 2416 different radial velocity bins toward 89 stars 
in the survey conducted by JT11, where all measurements of $f1$ and $f2$ with uncertainties 
less than 0.03 are depicted by dots whose areas are proportional to $N$(\ion{C}{1})$_{\rm 
total}$.  Here, we use colors to indicate local ultraviolet starlight densities, characterized by the 
parameter $\log I/I_0$, where $I_0$ is the average value in our part of the Galaxy (Mathis et al. 
1983).  Cases where the starlight density could not be determined are shown by x marks.  The 
curves indicate calculated outcomes for $f1$ and $f2$ for individual regions containing gas at 
three different temperatures, as indicated, and a range of thermal pressures indicated by dots 
spaced apart by 0.1 dex in $\log (p/k)$ and with numerals that signify whole-number values of 
$\log (p/k)~[{\rm cm}^{-3}$K]. The cross labeled  with an $\infty$ sign is at the location where 
the populations are in proportion to their statistical weights, representing very high densities 
and temperatures.\label{fig:JT11_plot}}
\end{figure}

The outcomes for the $f1$ and $f2$ results of JT11 are shown in Fig.~\ref{fig:JT11_plot}, where 
it is evident that nearly all of the points fall above the theoretical tracks that represent 
homogeneous regions at a single pressure. A small percentage of the sight lines showed 
substantially larger contributions from high pressure gas.  As noted by JT11,  there were two 
kinds of circumstantial evidence that favored the high pressures being related to stellar 
environments.  First,  the fractional amounts of high pressure gas seemed to be driven by the 
intensity of starlight  at the location of the gas, and second, material at predominantly negative 
velocities relative to expectations from differential Galactic rotation showed the best evidence 
of high pressures, presumably as a result of disturbances created by the target stars forcing the 
gas to move toward us.  The relationship to starlight density is clearly illustrated by the colors of 
points in Fig.~\ref{fig:JT11_plot}.  Many of the red points in that diagram indicate that strongly 
irradiated gas at certain velocities show at least half of the \ion{C}{1} in a high pressure 
environment.

Estimates for the column densities of neutral hydrogen $N$(H) along the sight lines in the JT11 
survey could be obtained from measures of $N$(\ion{O}{1}) or $N$(\ion{S}{2}), and 
characteristic volume densities $n$(H) arise from the measurements of thermal pressures, if 
the temperatures are known.  In turn, values of $N$(H)/$n$(H) yield an approximate 
longitudinal thickness of the \ion{C}{1}-bearing gas.  JT11 took such dimensions and divided 
them by the distances to the stars and found that most of the material occupied of order one 
percent of the length of the sight lines.  In most cases, it is likely that this gas could be located 
near the target star or its association of stars, probably because it is a remaining part of a dense 
cloud that collapsed and led to the star formation.  This consideration, along with the intensity 
and kinematic indications, suggests that most of the gas that was measured is situated in a 
region that might be subjected to disturbances such as mass loss, \ion{H}{2} region expansion, 
and different forms of radiation pressure from the star or its neighbors (Lamers 2001 ; Wareing 
et al. 2018 ; Barnes et al. 2020 ; Ali 2021).  Aside from these dynamical reasons for very high 
pressures, we may also consider that mild increases in pressure can arise when the enhanced 
radiation from stars can increase the level of heating by photoelectric emission from grains, 
which in turn establishes higher maximum and minimum pressures for the cold phase due to 
the upward shift for the thermal equilibrium curve in the representation of $\log (p/k)$ vs $\log 
n$(H) (Wolfire et al. 2003).

While the effects of the stars on their environments are of interest, we also wish to understand 
the nature of pressures in the general ISM that is well removed from the stars so that we can 
obtain better insights on the nature of turbulence in the general ISM.  If we could observe along 
sight lines that did not end at stars, would we still see evidence for gas at extraordinarily high 
pressures?  One incentive to explore this issue is a quest to see if the \ion{C}{1} results support 
findings on the existence of tiny scale atomic structures (TSAS) with extraordinarily high 
densities.  Stanimirović \& Zweibel (2018) have presented a review of various investigations on 
the nature of TSAS, most of which arise from observations of spatial and temporal variations of 
21-cm absorption (see also Rybarczyk et al. (2020)).  These observations have the shortcoming 
that the strengths of such absorptions depend on temperatures, which are often poorly known.  
For any given value of $N$(\ion{H}{1}) the \ion{C}{1} excitations increase with temperature, 
while the opposite is true for 21-cm absorption.  Fortunately, as one can see from the curves in 
Fig.~\ref{fig:JT11_plot}, the outcomes for pressures based on the \ion{C}{1} observations are 
not strongly dependent on temperature.

Small regions with unusually high pressures might arise from a manifestation of intermittency 
in turbulence, which can create shocks and provide very localized heating and compression of 
gas over a short time interval (Falgarone et al. 2015).  Large-scale colliding flows of warm gas 
can create an interface that contains highly compressed small cold clouds (Audit \& Hennebelle 
2005 ; Vázquez-Semadeni et al. 2006), whose internal pressures may be enhanced further by 
self gravity (Vázquez-Semadeni et al. 2007).  The more extreme cases may provide enough 
heating and ambipolar diffusion to facilitate the production of certain molecules that require 
endothermic reactions, such as CH$^+$, which is formed by the ion-neutral reaction C$^+ +{\rm 
H}_2\rightarrow {\rm CH}^++{\rm H}$ ($\Delta E/k=-4640$\,K) (Draine \& Katz 1986 ; Myers et 
al. 2015).

As a supplement to existing data on TSAS and observations of turbulent intermittency, we can 
examine the \ion{C}{1} excitations, but we must use spectra from extragalactic sources of UV 
radiation to avoid the effects of stars.  Several examples have already been published  for sight 
lines toward the Magellanic Clouds (Nasoudi-Shoar et al. 2010 ; Welty et al. 2016). Welty et al. 
(2016) found the average values $f1=0.21$ and $f2=0.07$ for the Galactic velocity components 
in the spectra of four stars in the Magellanic Clouds, and these values are almost exactly equal 
to the average values found by JT11.

\section{Measurements of \ion{C}{1} toward Extragalactic Targets}\label{sec:extragalactic}

Extragalactic sources are considerably fainter than the stars utilized in the JT11 survey.  
Nevertheless, we have identified a few cases where intensive observations yielded spectra of 
sufficiently good quality to warrant investigations of the \ion{C}{1} features.  All observations 
reported here are based on spectra recorded with the medium resolution E140M echelle mode 
of STIS on HST.\footnote{Specific details on the dates of the observations, exposure times, and 
central wavelength settings for all observations that we used in this paper can be accessed via 
the following doi for data in the {\it Mikulsky Archive for Space Telescopes\/} ({\it MAST\/}): 
\dataset[10.17909/t9-e05b-1j72]{\doi{10.17909/t9-e05b-1j72}}.}

The Galactic coordinates, V magnitudes, and relevant velocities for \ion{C}{1} for the selected 
targets are listed in Table~\ref{tbl:targets}.   We must acknowledge that there is a selection bias 
toward sight lines that exhibited enough \ion{C}{1} to measure above the noise level with some 
reliability.  Nevertheless, our sight lines typically sample less gas than the determinations 
carried out in the survey by JT11, which had a median hydrogen column density of about 
$2\times 10^{21}{\rm cm}^{-2}$.  For our Galactic latitudes that range from $23\arcdeg$ to 
$64\arcdeg$, we anticipate that if we use a model for the vertical distribution of \ion{H}{1} 
expressed by McKee et al. (2015), our sight lines probably sample an approximate range $0.45 
\lt N$(\ion{H}{1})$ \lt 1.9\times 10^{21}{\rm cm}^{-2}$.

\begin{figure}
\epsscale{1.3}
\plotone{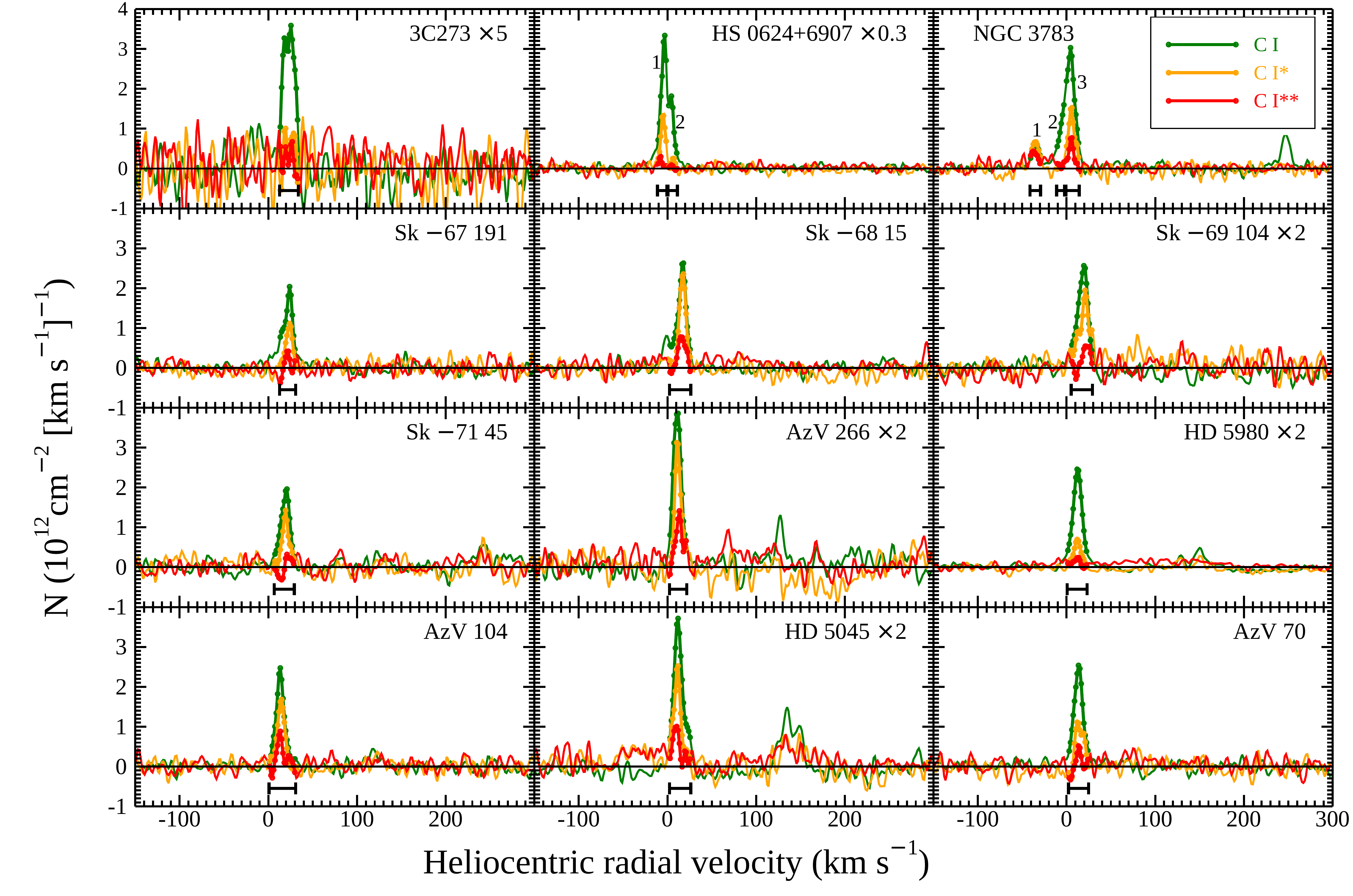}
\caption{Derived values of column densities per unit velocity interval for \ion{C}{1} in the three 
fine-structure levels as a function of heliocentric radial velocity, with colors matched to the 
respective levels as indicated.  Displays for some targets had their amplitudes changed by 
uniform factors, as indicated near their identifications, to show the profiles more clearly on the 
uniform $y$-axis scales.  Velocity intervals where the profiles were sampled are indicated by 
bold lines with dots on the profiles themselves and range bars at the bottom of each plot.  For 
some targets, there were two or more such samples whose locations are indicated by numbers, 
and these samples are identified in separate panels in Fig.~\protect\ref{fig:prob_plots1}.  
Samples shown in Fig~\protect\ref{fig:prob_plots2} had only single velocity ranges of interest.  
For sight lines toward the Magellanic Clouds, \ion{C}{1} peaks that occasionally appear at $v\gt 
100\kms$ arise from gas within the clouds, which is not relevant to our investigation.  The peak 
at +250\kms for NGC~3783 is produced by material in the high velocity cloud (HVC) 
287.5+22.5+240 (aka WW187) in the leading arm of the Magellanic Stream (West et al. 1985 ; 
Lu et al. 1998 ; Sembach et al. 2001 ; Wakker 2001). \label{fig:ci_profiles}}
\end{figure}

 Profiles of column densities vs. velocity for the three \ion{C}{1} levels for each sight line are 
shown in Fig.~\ref{fig:ci_profiles}.  This figure shows the velocity intervals over which we made 
our measurements, and these ranges are listed in Column~5 of Table~\ref{tbl:targets}.  For the 
targets in the Magellanic Clouds, we have intentionally excluded the velocity ranges for 
absorption in these systems, since our objective is to measure \ion{C}{1} that is not in the 
vicinity of the target stars. 

\begin{deluxetable}{
r	
c	
c	
c	
c	
c	
c	
c	
}
\tablewidth{0pt}
\tablecolumns{8}
\tablecaption{Targets and their \ion{C}{1} Results\label{tbl:targets}}
\tablehead{
\colhead{} & \colhead{V} & \multicolumn{2}{c}{Gal. Coord.} & \colhead{Velocity 
Range\tablenotemark{a}} & \multicolumn{3}{c}{Column Densities ($10^{12}{\rm cm}^{-
2}[\kms]^{-1})$\tablenotemark{b}}\\
\cline{3-4} \cline{6-8}
\colhead{Target} & \colhead{mag.} & \colhead{$\ell$} & \colhead{$b$} & \colhead{\kms 
(heliocentric)} & \colhead{$N$(\ion{C}{1})} & \colhead{$N$(\ion{C}{1}$^*$)} & 
\colhead{$N$(\ion{C}{1}$^{**}$)}\\
\colhead{(1)} & \colhead{(2)} & \colhead{(3)} & \colhead{(4)} & \colhead{(5)} & \colhead{(6)} & 
\colhead{(7)} & \colhead{(8)}
}
\startdata
\cutinhead{Distant Objects}
\object{3C273}&14.83&289.95&+64.36&12.75 to 33.75 &$ 0.62\pm 0.05$&$ 0.14\pm 
0.08$&$ 0.07\pm 0.07$\\
\object{HS 0624+6907}&14.16&145.71&+23.35&$-11.25$ to $-0.75$ &$ 7.50\pm 0.56$&$ 
2.97\pm 0.25$&$ 0.52\pm 0.25$\\
&&&&0.75 to 11.25 &$ 4.70\pm 0.39$&$ 0.50\pm 0.23$&$ 0.13\pm 0.24$\\
\object{NGC 3783}&13.43&287.45&+22.95&$-41.25$ to $-29.25$ &$ 0.49\pm 0.04$&$ 
0.57\pm 0.10$&$ 0.35\pm 0.10$\\
&&&&$-11.25$ to  $-2.25$ &$ 1.26\pm 0.05$&$ 0.09\pm 0.11$&$ 0.09\pm 0.11$\\
&&&&$-0.75$ to 14.25 &$ 2.51\pm 0.05$&$ 1.15\pm 0.11$&$ 0.51\pm 0.10$\\
\cutinhead{Magellanic Cloud Stars}
\object{Sk $-$67 191}&13.44&277.66&$-$32.33&12.75 to 30.75 &$ 1.59\pm 0.09$&$ 0.85\pm 
0.11$&$ 0.19\pm 0.10$\\
\object{Sk $-$68 15}&12.69&279.52&$-$35.47&2.25 to 26.25 &$ 2.10\pm 0.08$&$ 1.83\pm 
0.16$&$ 0.63\pm 0.13$\\
\object{Sk $-$69 104}&12.10&279.92&$-$33.39&5.25 to 29.25 &$ 1.01\pm 0.07$&$ 0.70\pm 
0.10$&$ 0.18\pm 0.07$\\
\object{Sk $-$71 45}&11.55&281.86&$-$32.02&6.75 to 29.25 &$ 1.60\pm 0.17$&$ 1.01\pm 
0.16$&$ 0.12\pm 0.13$\\
\object{AzV 266}&12.61&301.90&$-$44.65&2.25 to 21.75 &$ 1.63\pm 0.10$&$ 1.16\pm 
0.14$&$ 0.47\pm 0.09$\\
\object{HD 5980}&11.31&302.07&$-$44.95&0.75 to 23.25 &$ 1.01\pm 0.03$&$ 0.26\pm 
0.04$&$ 0.08\pm 0.02$\\
\object{AzV 104}&13.22&302.91&$-$44.33&2.25 to 30.00 &$ 1.94\pm 0.08$&$ 1.32\pm 
0.11$&$ 0.58\pm 0.10$\\
\object{HD 5045}&11.04&303.01&$-$43.66&2.25 to 26.25 &$ 1.46\pm 0.08$&$ 0.92\pm 
0.10$&$ 0.36\pm 0.09$\\
\object{AzV 70}&12.40&303.05&$-$44.49&2.25 to 24.75 &$ 2.08\pm 0.09$&$ 0.90\pm 
0.14$&$ 0.28\pm 0.13$\\
\enddata
\tablenotetext{a}{Range over which \ion{C}{1}, \ion{C}{1}$^*$, and \ion{C}{1}$^{**}$ were 
sampled.  Targets with multiple ranges are identified with component numbers that are 
ordered in accord with the velocity midpoints.  We have excluded velocity ranges attributable 
to the Magellanic Clouds and the leading arm of the Magellanic Stream.}
\tablenotetext{b}{Weighted average values over the velocity range specified in Column~5.}
\end{deluxetable}

In our analysis of the profiles of \ion{C}{1}, \ion{C}{1}$^*$, and \ion{C}{1}$^{**}$, we had to 
acknowledge the existence of two types of errors.  First, there are noise fluctuations that arise 
from photon counting, and these errors impact the solutions for the column densities within 
every velocity bin.  These errors become worse when the optical depths are large, which 
happens for some cases.  In other cases, the lines are weak, and such features are affected by 
another source of uncertainty that is also significant.  When continuum levels are not accurately 
determined, which is often the case for stellar targets that have very broad features in their 
spectra, there can be erroneous offsets in column densities that are sustained over very long 
intervals in velocity.  These offsets are clearly evident in the tracings shown in 
Fig.~\ref{fig:ci_profiles}, and we interpret them quantitatively by measuring them at velocities 
well removed from the \ion{C}{1} absorptions.

We determine column densities for the three \ion{C}{1} states from measurements that span 
many adjacent velocity bins where the contributions are visible, as indicated in 
Fig.~\ref{fig:ci_profiles}.  For the random errors arising from photon noise, we can consider that 
they are independent of each other from one velocity bin to the next.\footnote{The separation 
of velocity bins is one-half of the STIS pixel pitch, so the errors in adjacent bins are correlated.  
We compensate for this when we sum over many bins be increasing the apparent errors by 
$\sqrt{2}$.}  However, the more sustained errors from misplaced continua (i.e., long 
wavelength undulations) are mostly coherent over an entire measurement interval.  Thus, 
these baseline errors can cause erroneous uniform shifts in the column densities across the 
complete profiles.  For the observations of stellar targets, these shifts typically increased the 
uncertainties by a factor of two over those attributable to noise alone.  For the distant objects 
consisting of active galactic nuclei, the spectra are more uniform with wavelength, and hence 
continuum errors are much smaller.

Another feature of our analysis is that when we average the column density results across a 
profile, we apply weight factors for the measurements in the velocity bins that are proportional 
$\{N$(\ion{C}{1}$)_{\rm total}/\sigma[N$(\ion{C}{1}$)_{\rm total}]\}^2$, where 
$\sigma[N$(\ion{C}{1}$)_{\rm total}]$ is the uncertainty in $N$(\ion{C}{1}$)_{\rm total}$.  This 
weighting scheme emphasizes results near the peaks of the absorptions, which are more 
reliable than those near the lower amplitude wings.  These weighted average column densities 
per unit velocity interval, along with their uncertainties, are listed in Columns~6$-$8 of 
Table~\ref{tbl:targets}.

For all of our cases, our objective is not only to determine the best values of $f1$ and $f2$, but 
also their uncertainties, given our expected measurement errors.  The nature of these level 
fractions is such that the uncertainties, when large, cannot be expressed in the form of simple 
numerical quantities that are independent of each other.  Differential probabilities in the
$f1$--$f2$ space can manifest themselves in unusual shapes.  In order to understand better the 
range of acceptable combinations of $f1$ and $f2$, we performed Markov Chain Monte Carlo 
(MCMC) calculations with Gibbs sampling that made successive comparisons of the $\chi^2$ 
outcomes for random trial values of $f1$, $f2$, and $N$(\ion{C}{1})$_{\rm total}$.  This 
calculation created probability densities for $f1$ and $f2$ marginalized over the quantity 
$N$(\ion{C}{1})$_{\rm total}$, but we restricted combinations of these two variables to a 
region in their diagram that is physically possible for mixtures of gases at different pressures.  
This region is bounded by the theoretical curves on the right-hand side and a line on the
left-hand side that spans between $(f1,f2)=(0,0)$ and fractions appropriate to very high 
temperatures and densities $(f1,f2)=(1/3,5/9)$, labeled $\infty$ in Figs.~\ref{fig:JT11_plot},
\ref{fig:prob_plots1}, and \ref{fig:prob_plots2}.

\section{Outcomes}\label{sec:outcomes}

\begin{figure}
\gridline{\fig{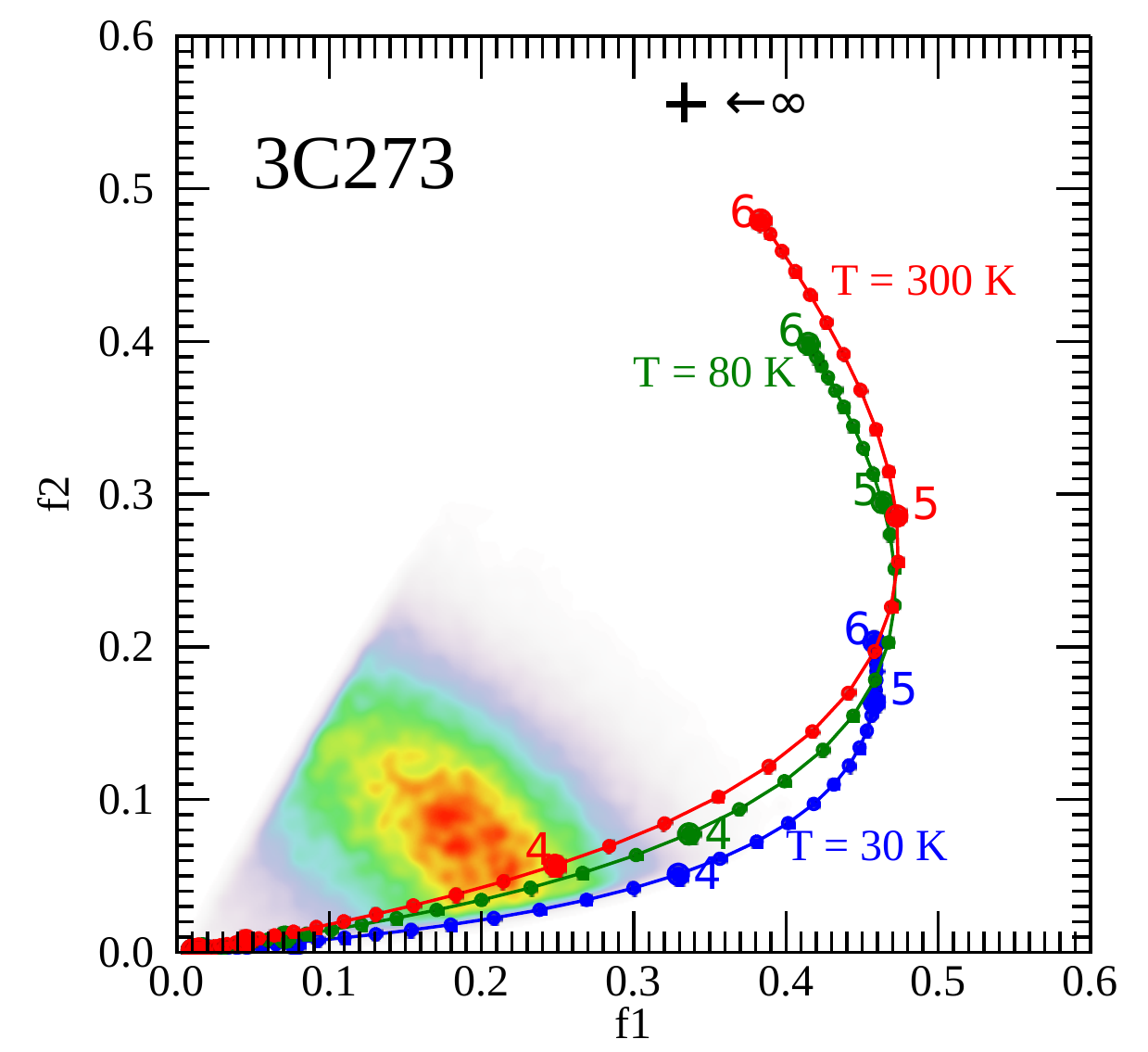}{0.3\textwidth}{(a)}
               \fig{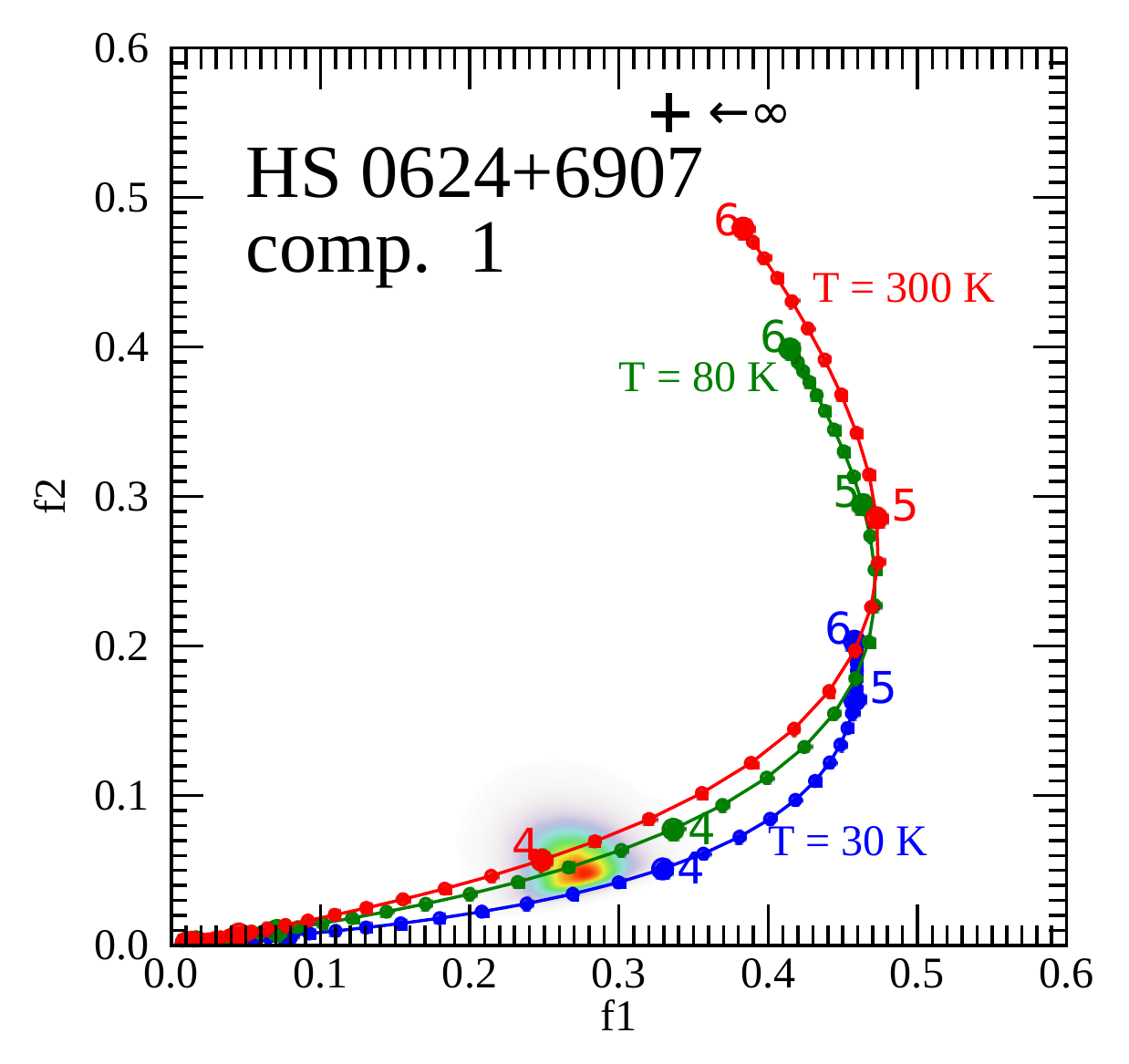}{0.3\textwidth}{(b)}
               \fig{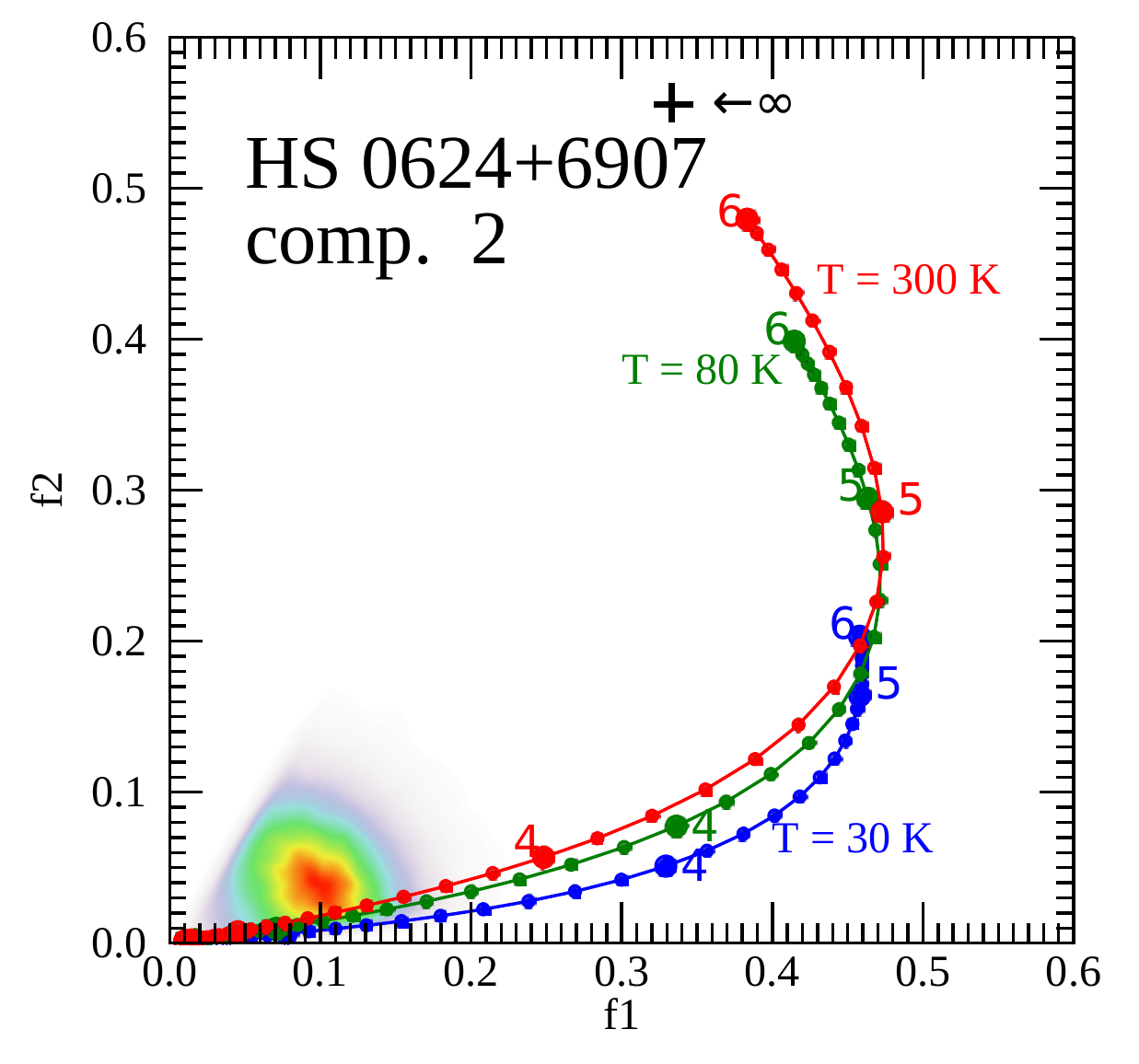}{0.3\textwidth}{(c)}}
\gridline{\fig{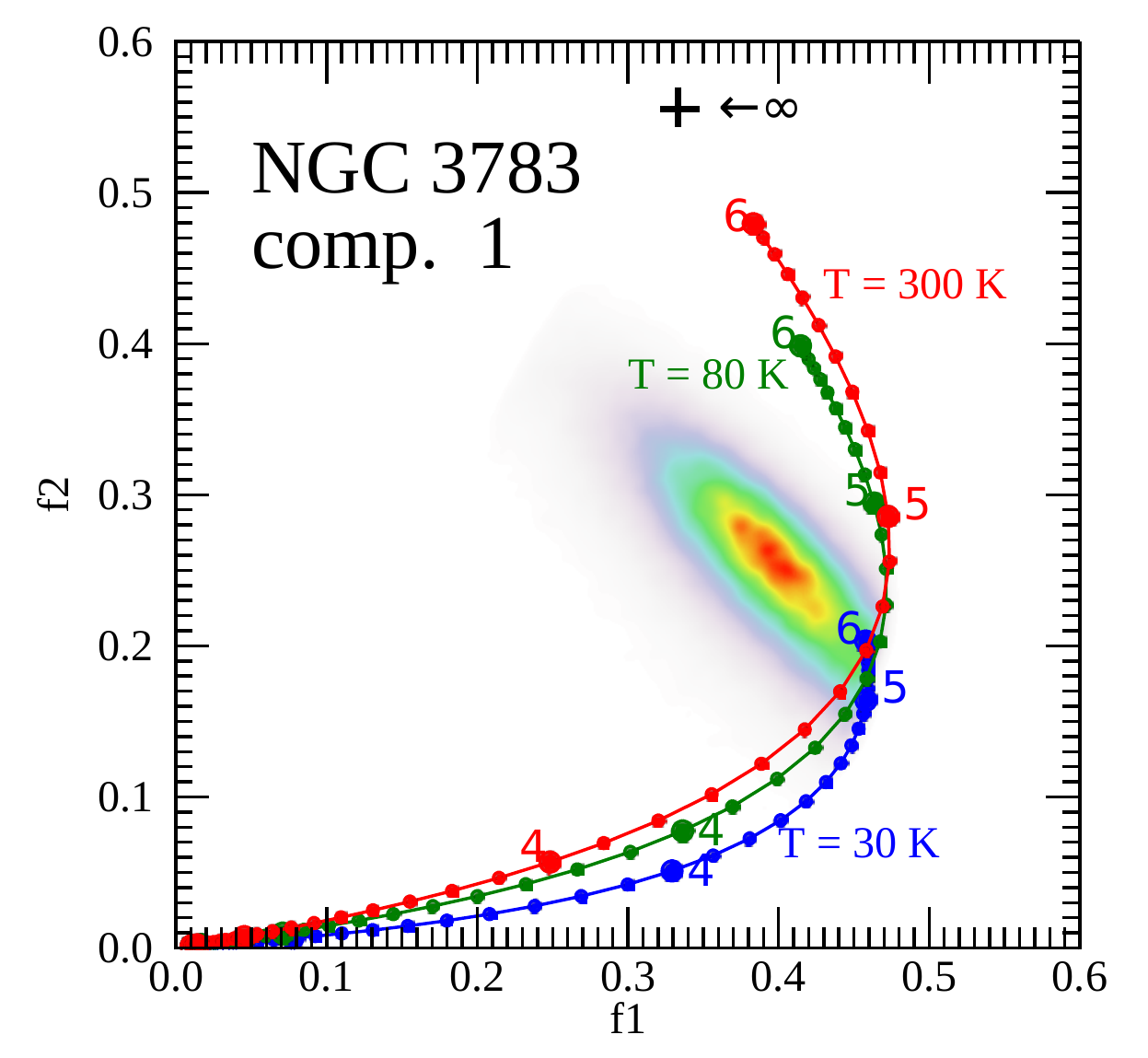}{0.3\textwidth}{(d)}
                \fig{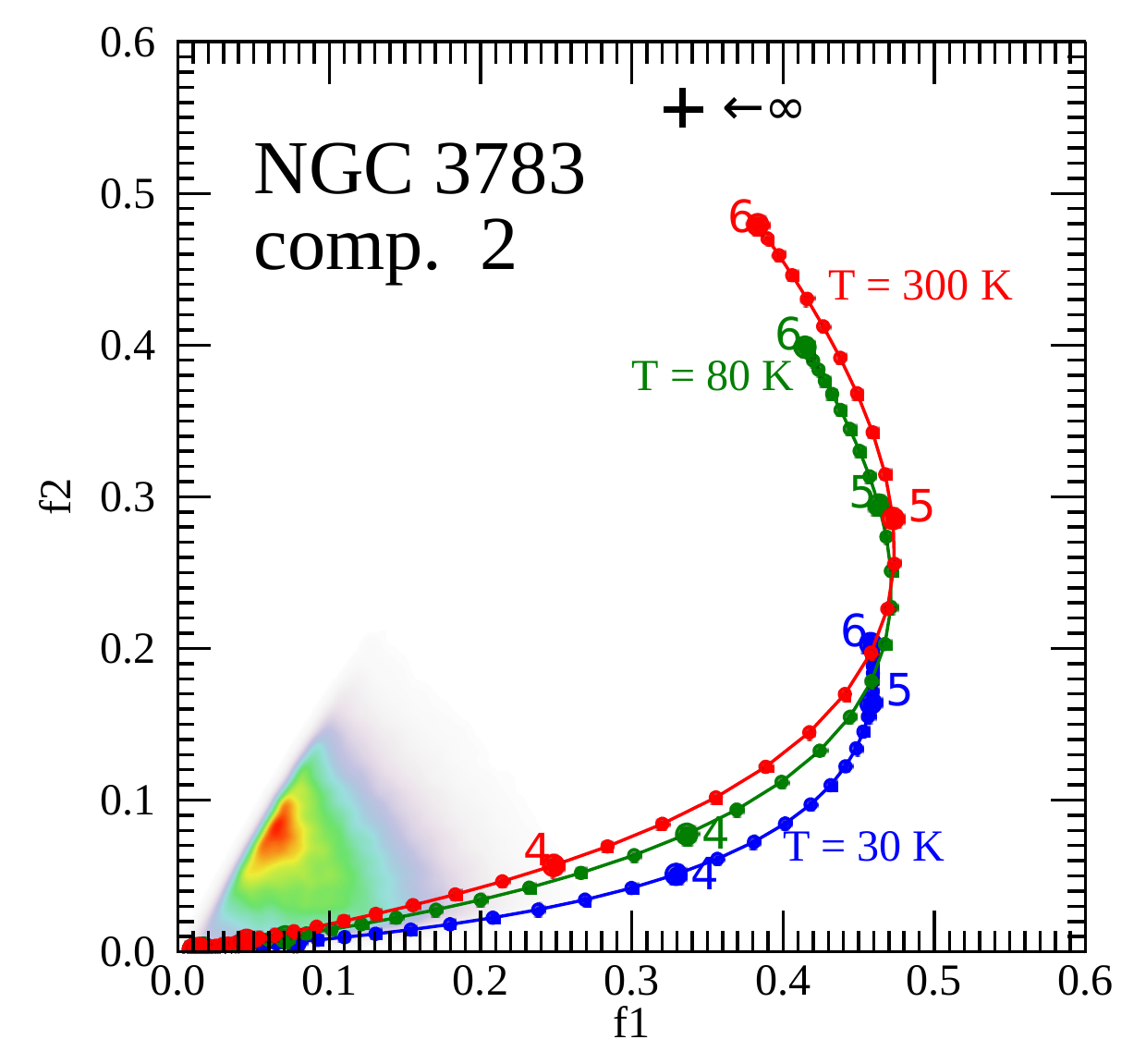}{0.3\textwidth}{(e)}
                \fig{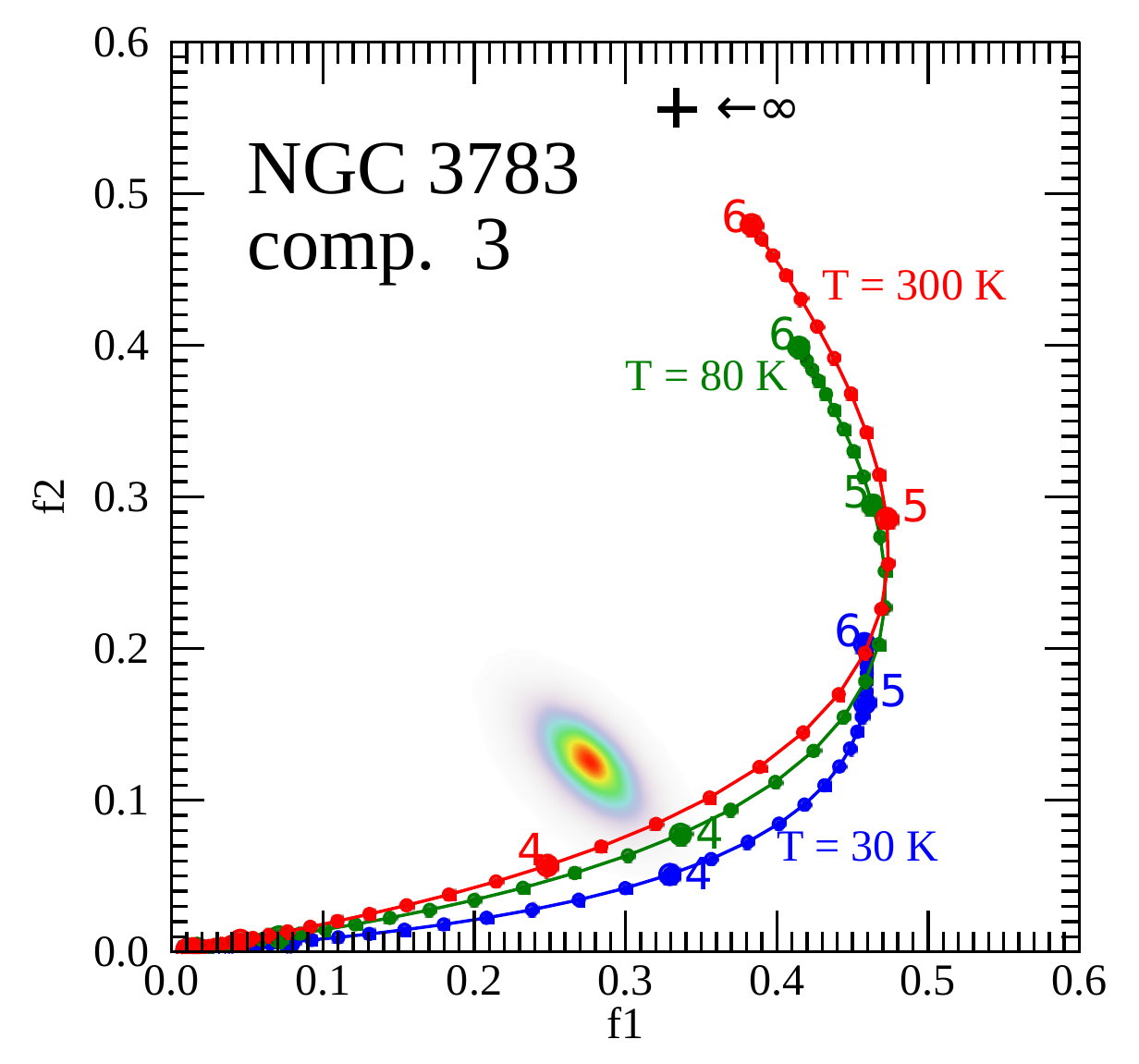}{0.3\textwidth}{(f)}}
\caption{Probability densities for $f1$ and $f2$ for \ion{C}{1} Galactic absorption toward (a) 
3C273, and (b,c) two parts of the velocity component toward HS0624+6907.  Panels (d,e,f) 
show three velocity intervals for the results toward NGC3783.\label{fig:prob_plots1}}
\end{figure}
\begin{figure}
\gridline{\fig{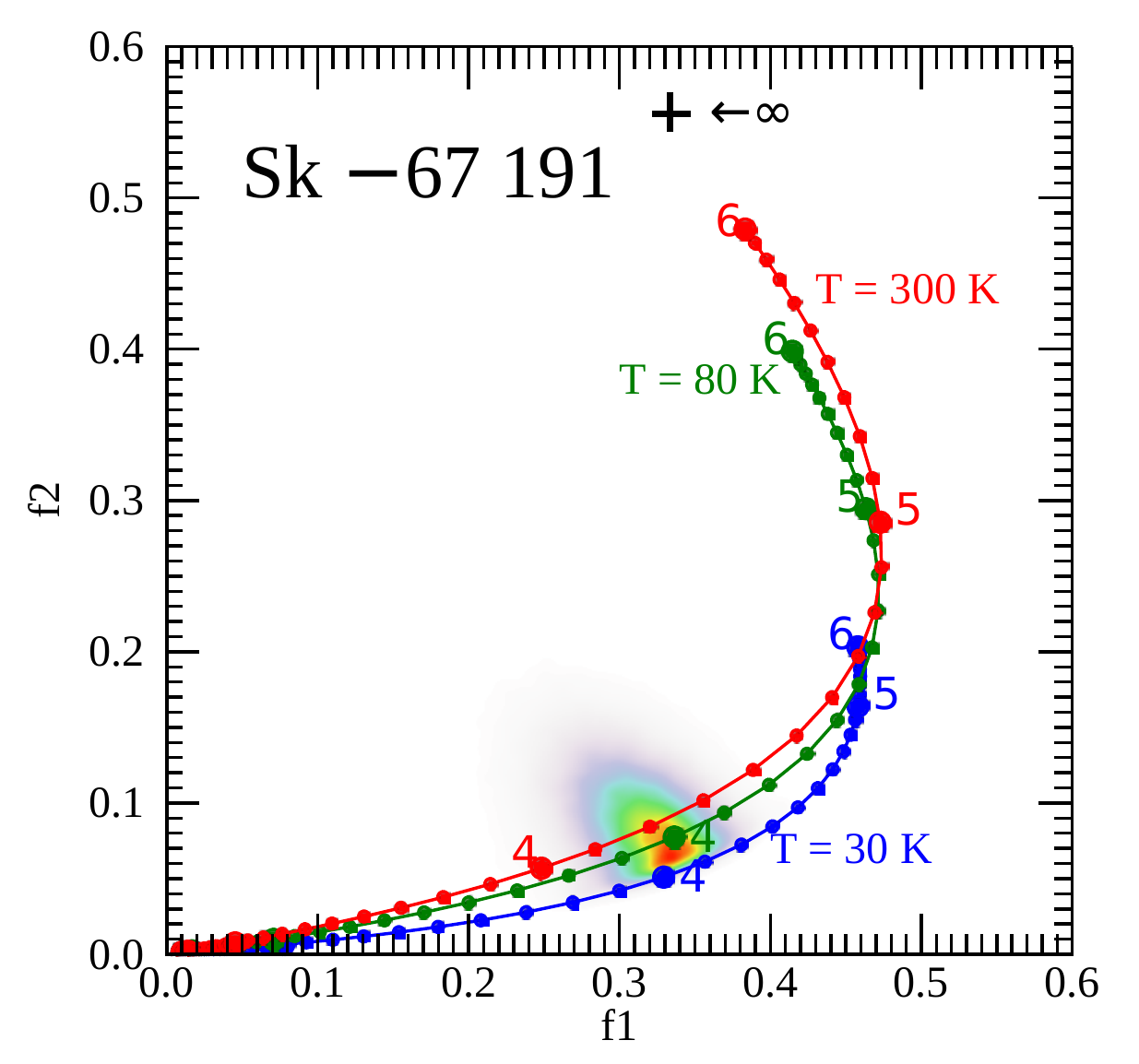}{0.3\textwidth}{(a)}
               \fig{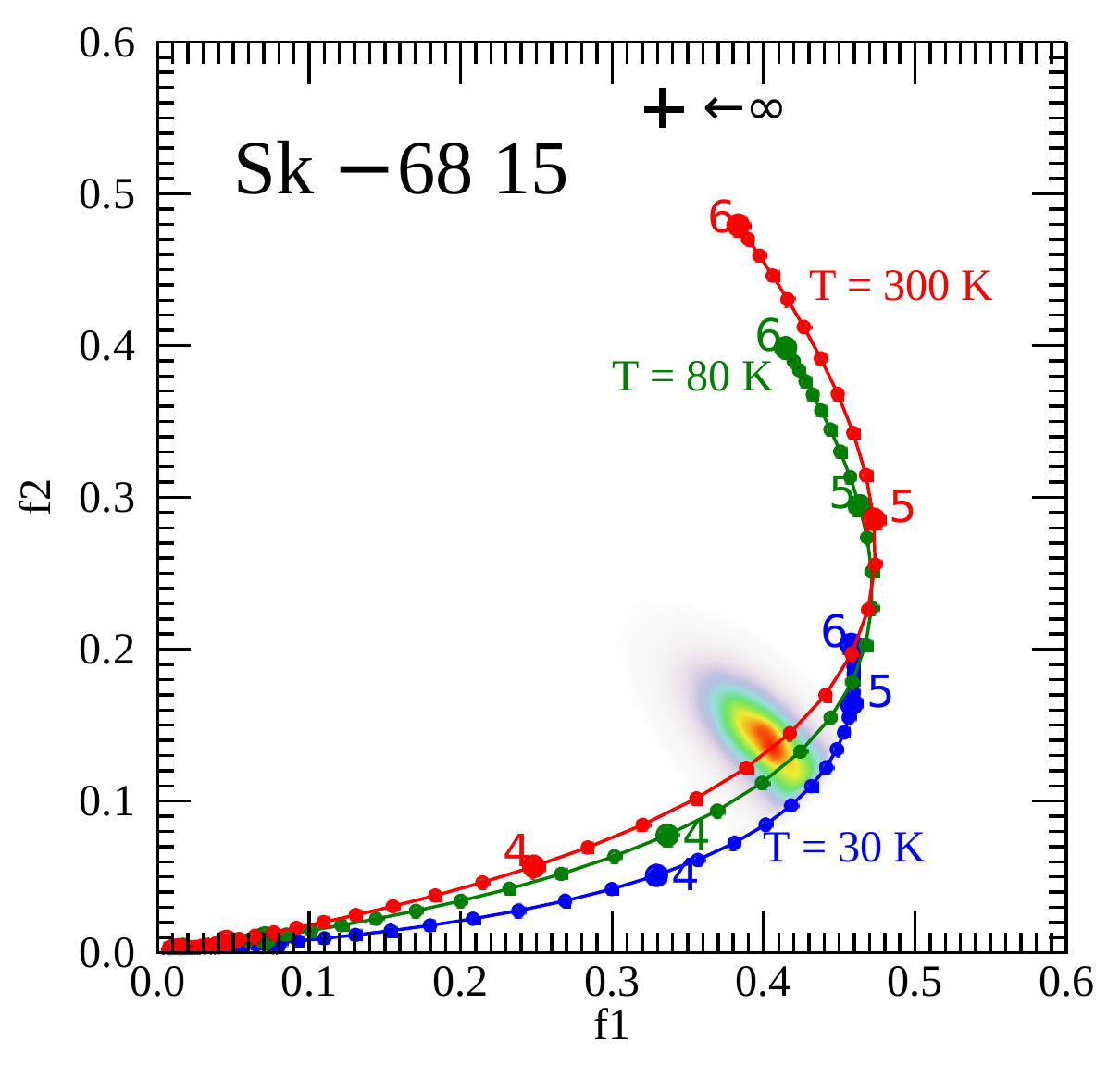}{0.293\textwidth}{(b)}
               \fig{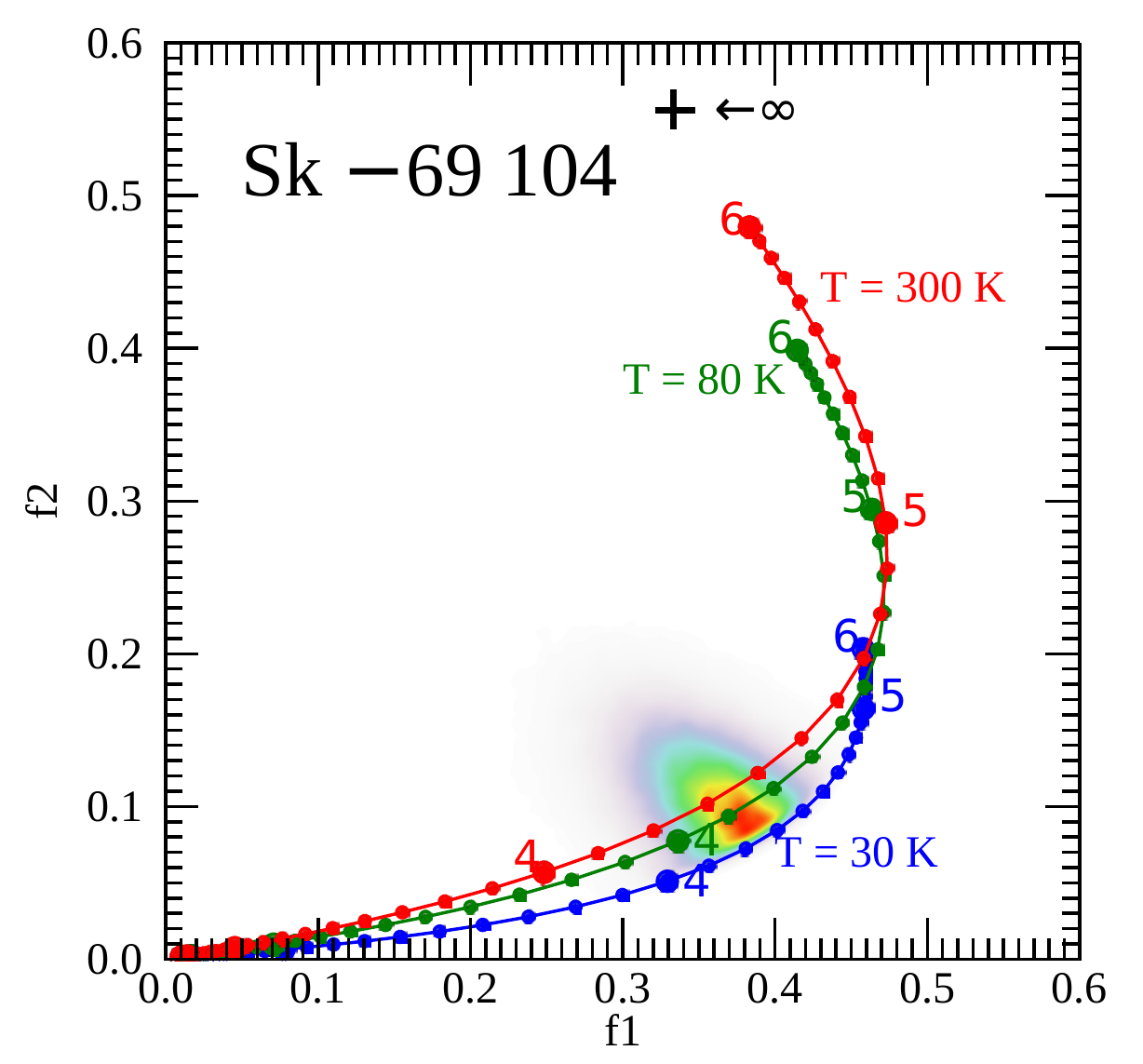}{0.3\textwidth}{(c)}}
\gridline{\fig{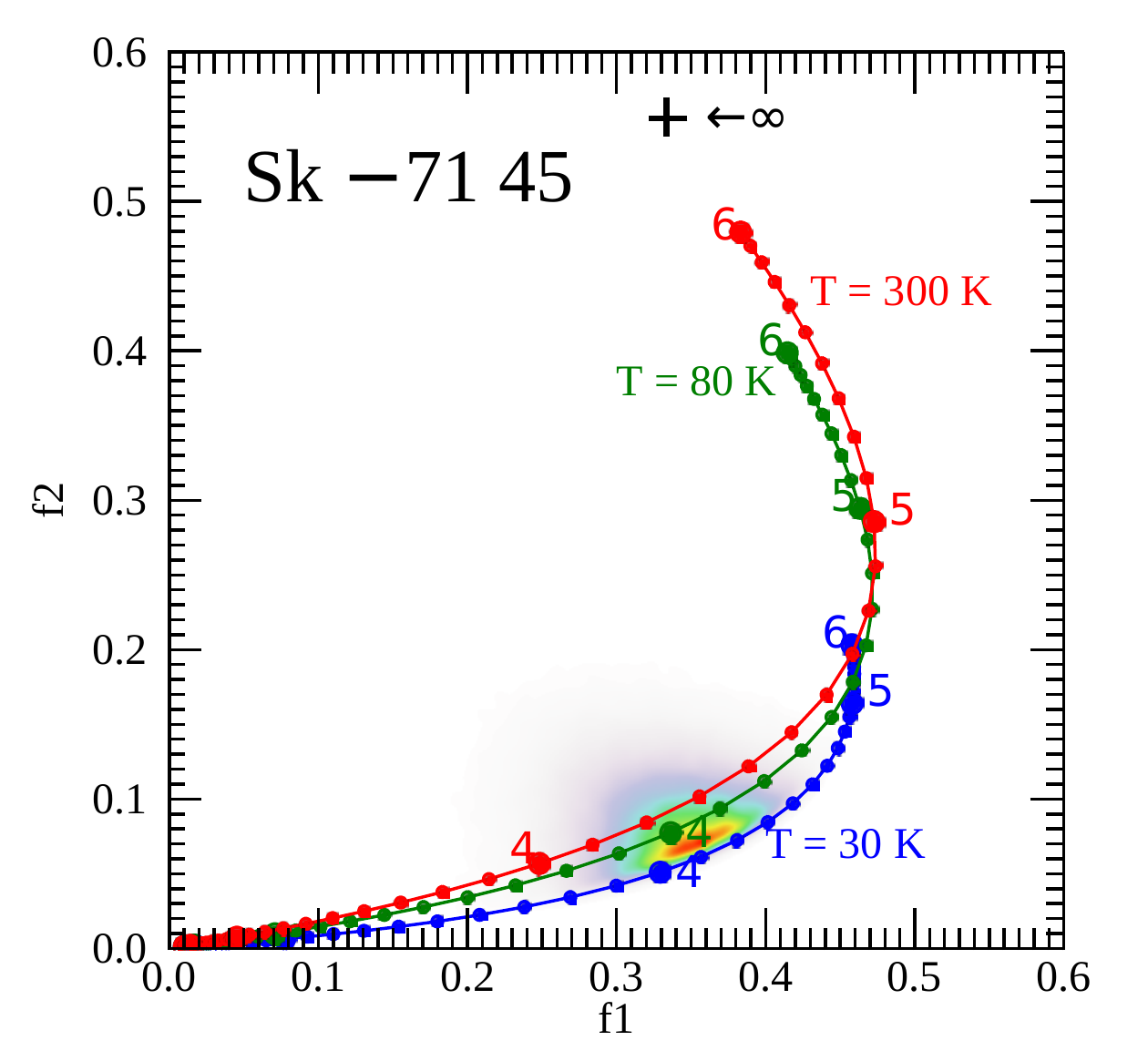}{0.3\textwidth}{(d)}
                \fig{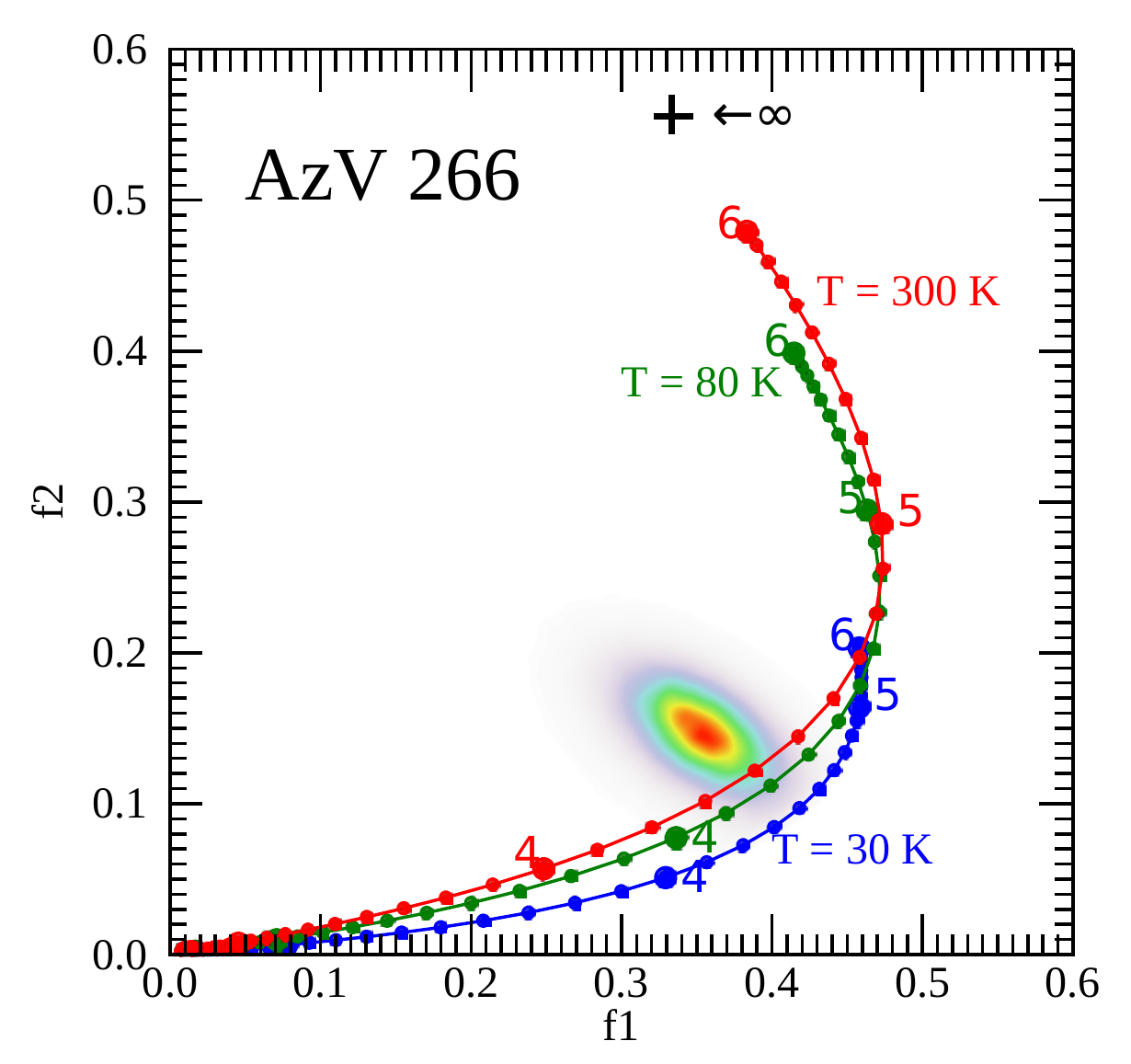}{0.3\textwidth}{(e)}
                \fig{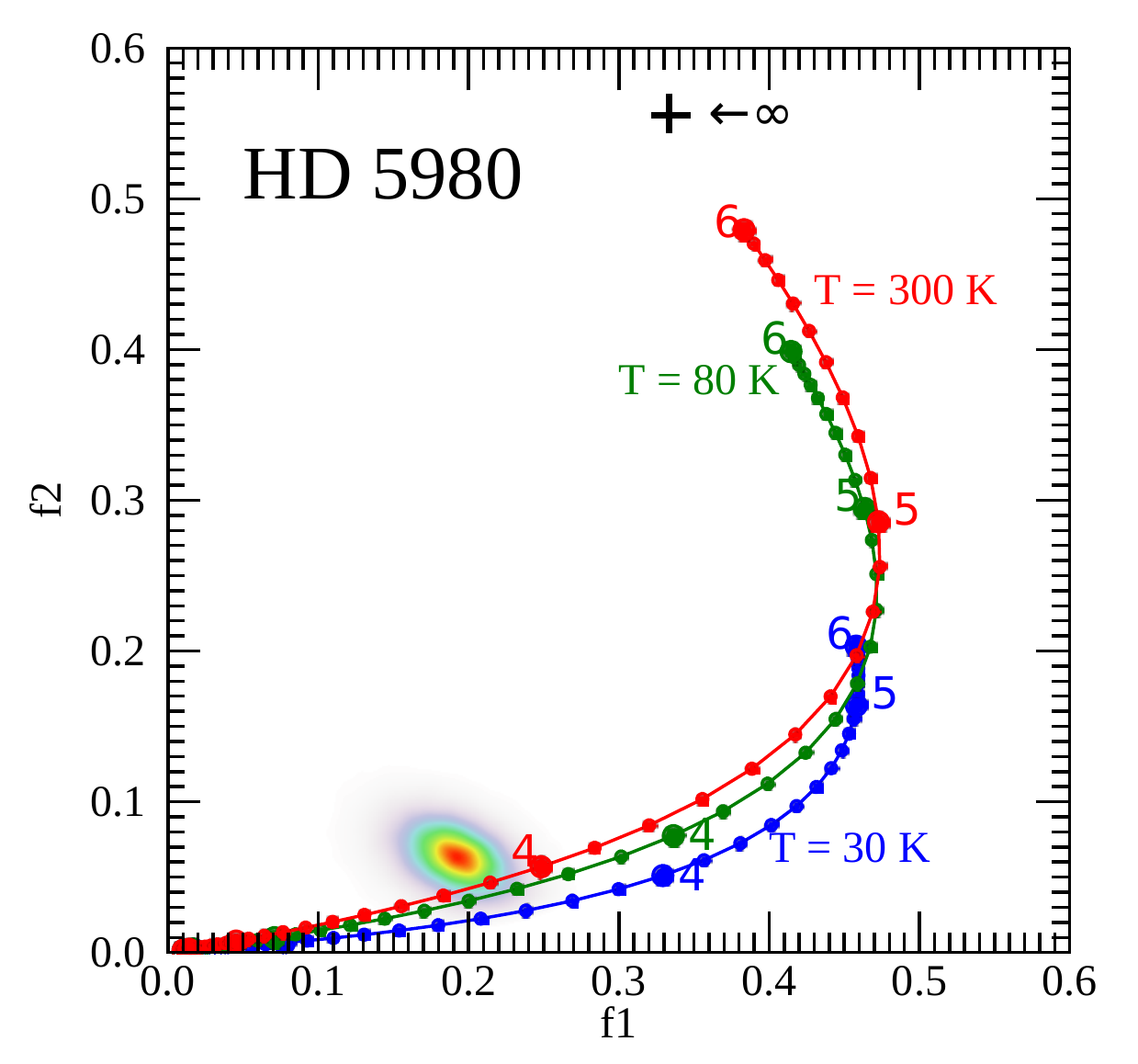}{0.3\textwidth}{(f)}}
\gridline{\fig{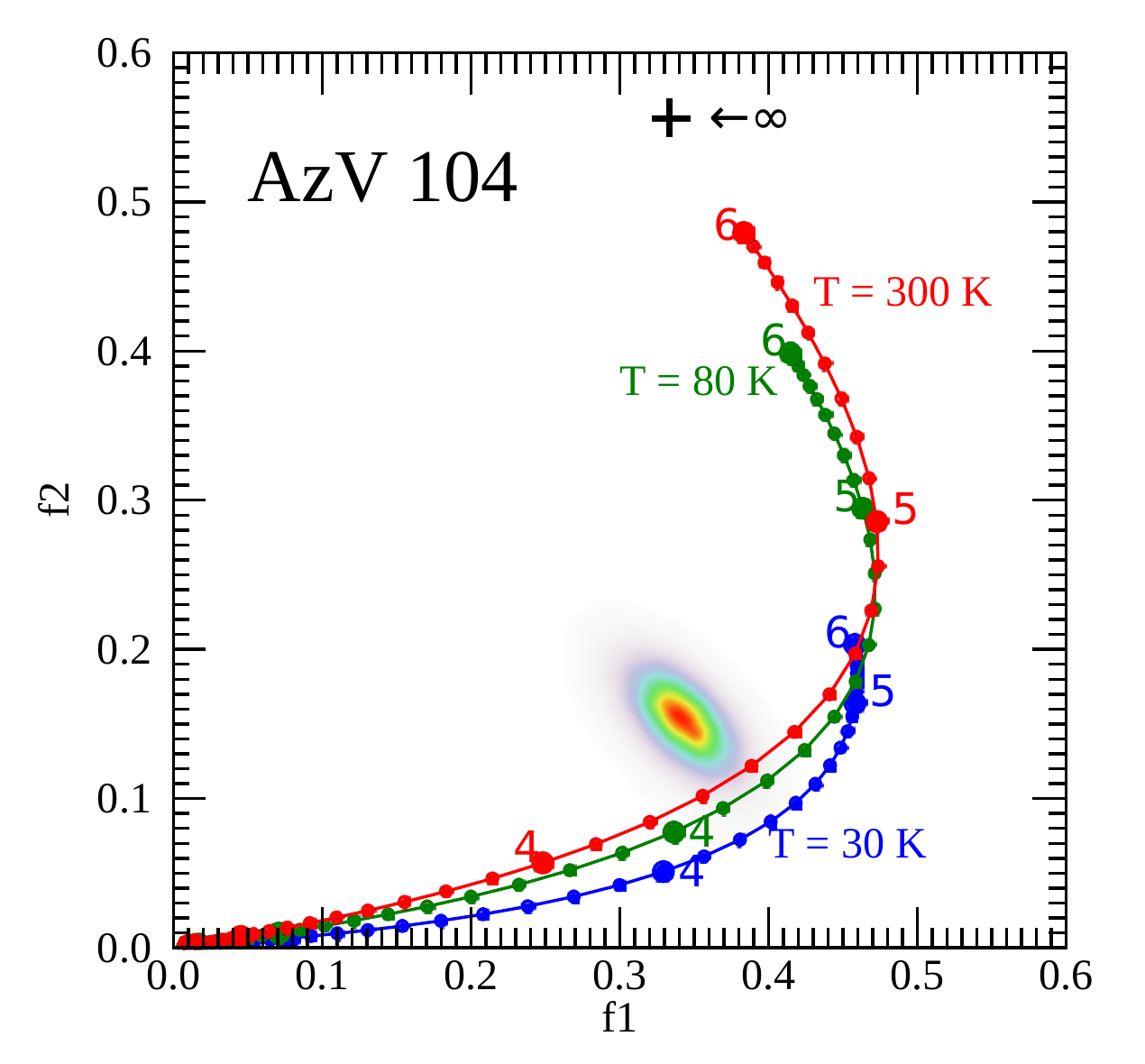}{0.3\textwidth}{(g)}
                \fig{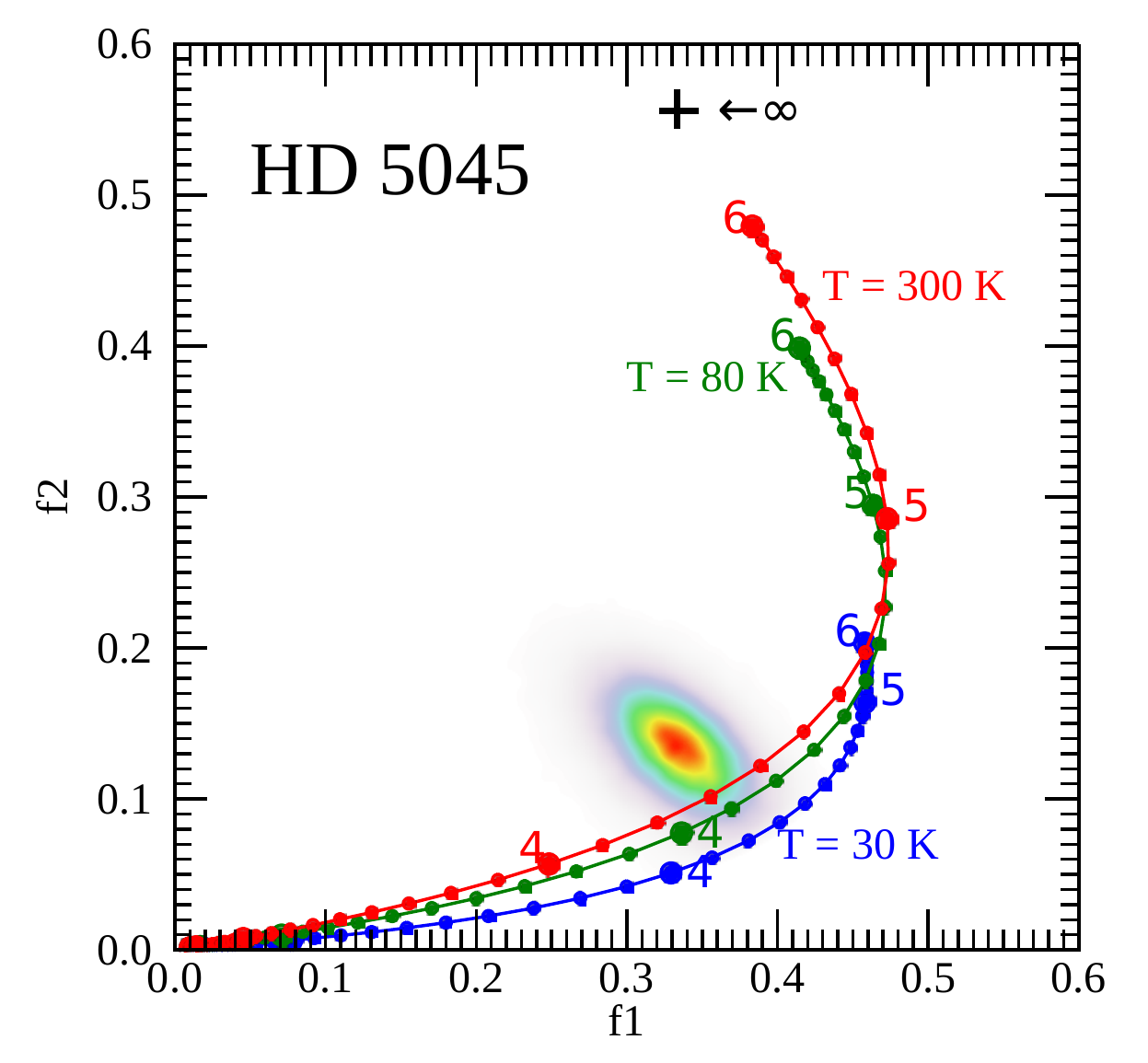}{0.3\textwidth}{(h)}
                \fig{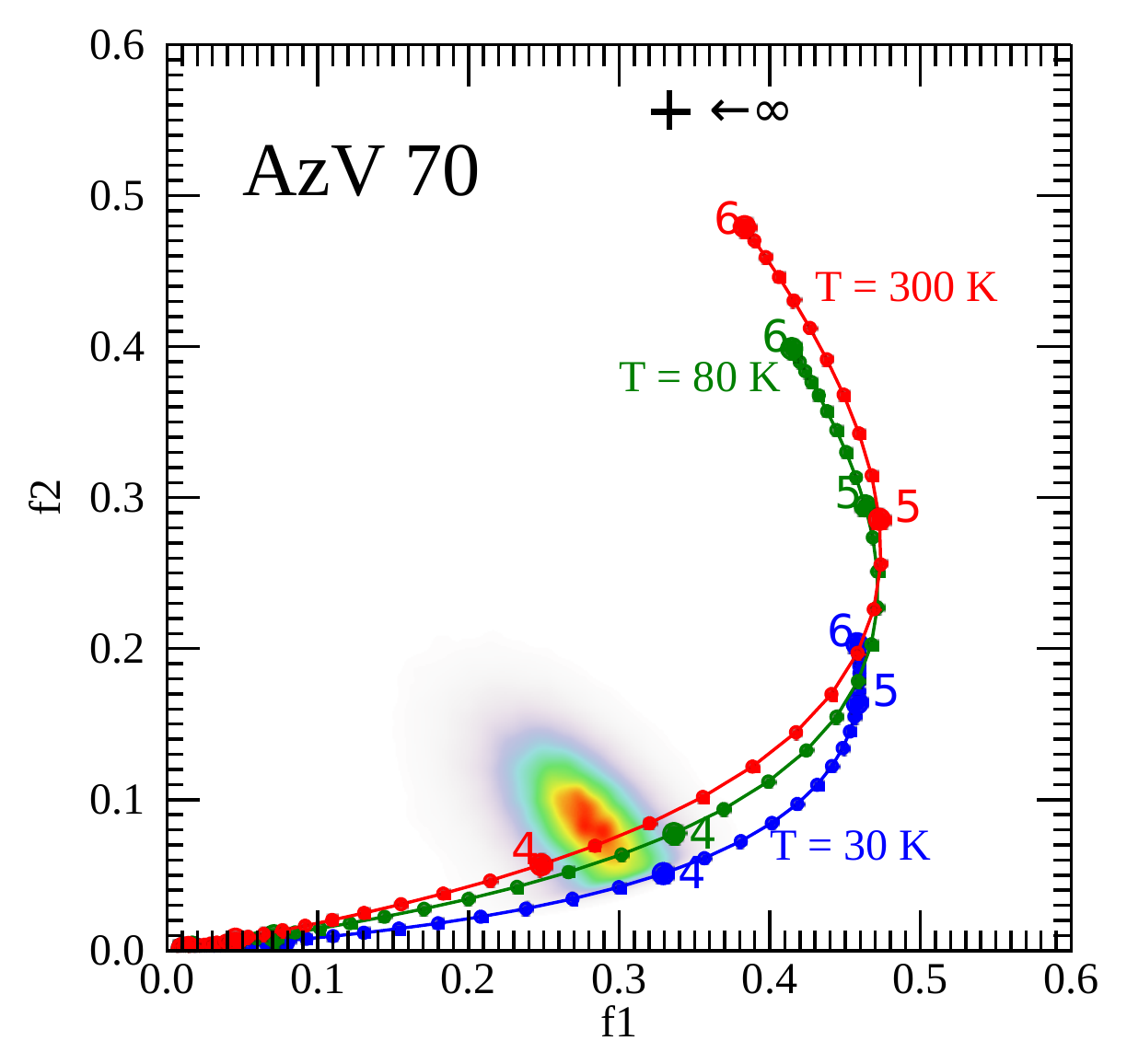}{0.3\textwidth}{(i)}}

\caption{Probability densities for $f1$ and $f2$ for \ion{C}{1} Galactic absorption toward stars 
in the LMC (panels a-d) and SMC (panels e-i).\label{fig:prob_plots2}}
\end{figure}

In this study, we have only 12 sight lines, so we do not have the vast array of measurements like 
those shown in Fig.~\ref{fig:JT11_plot} that were obtained by JT11, which allowed them to 
construct a pressure distribution function for their low pressure component.  Instead, the 
current sample, while small, liberates us from the influence of material surrounding hot stars 
(although not completely, as we will describe later).  The questions that we wish to answer are 
(1) does the distribution function of low pressure material differ from the result obtained by 
JT11, and (2) whether or not the small amounts of very high pressure gas seen in the JT11 
survey are no longer evident or at least substantially diminished.

Figures~\ref{fig:prob_plots1} and \ref{fig:prob_plots2} show the probability densities for $f1$ 
and $f2$ obtained from the MCMC analyses for our targets.  The location on the diagram for 
Component~1 centered on $-35\kms$ for NGC~3783 stands out markedly from the other 
determinations by showing a substantial fraction of the \ion{C}{1} is at $p/k=10^5\cmk$.  Since 
this result clearly contrasts with the others, we will treat it as an extraordinary situation and will 
examine it in more detail in the following section (\ref{sec:NGC3783}). 
\newpage
\section{Interpretation}\label{sec:interpretation}

\subsection{The $\mathit{-35~km~s^{-1}}$ Component toward NGC 3783}\label{sec:NGC3783}

As we indicated earlier, the isolated Component~1 centered at a velocity of $-35\kms$ toward 
NGC~3783 showed that a substantial portion of the \ion{C}{1} is composed of gas with a 
pressure of about $p/k=10^5\cmk$ (see the lower leftmost panel in Fig.~\ref{fig:prob_plots1}).  
However, we subsequently discovered that this anomalous outcome could be the result of two 
unusual conditions for this particular sight line.  First, it is nearly aligned with a hot, luminous 
star (HD~101274, spectral type A0V, V magnitude = 9.12) with a separation of only 72\farcs9 
from our target.  This star has a Gaia EDR3 listed parallax of $2.13\pm 0.02$~mas, so the sight 
line passes to within about 0.166~pc of the star.  The contrast between the \ion{C}{1} for this 
component compared to all of the other measurements may be a confirmation of our 
proposition that stellar environments harbor high pressure gas.  It is possible that 
Component~3 centered on 8\kms toward this target may also be influenced by the star, since it 
is one of only a few cases that clearly show evidence of a small amount of high pressure 
material.

We estimate that from the star’s spectral type that it should have an effective temperature 
$T=10,000$\,K and a radius of $6.76R_\odot$.  From the stellar flux at this temperature 
modeled by Fossati et al. (2018), we calculate that we would need to have $n_e\lt 0.06\,{\rm 
cm}^{-3}$ to obtain the radius of a Strömgren Sphere that extends beyond the 0.166~pc impact 
parameter.  In our part of the Galaxy, an electron density $n_e\approx 0.04\,{\rm cm}^{-3}$ is a 
reasonable outcome for $n$(H)$=0.5\,{\rm cm}^{-3}$ (Jenkins 2013). Another possibility is that 
we are sensing material that has been compressed in a bow shock or thick shell caused by mass 
loss from the star.  While mass loss rates from A0~V stars are not nearly as large as those from 
hotter, more  luminous O- and B-type stars, they may still exhibit a strong enough outflow to 
create an over-pressurized shell of neutral gas
(Castor et al. 1975 ; Weaver et al. 1977 ; Lancaster et al. 2021a,b).  Following the finding 
of Beaumont et al. (2014) that such shells can be detected by their infrared emission, we 
examined archival {\it Spitzer\/} IRAC images of HD~101274 exposed at 3.6, 4.5, and $8\mu$m, 
but we could not see any evidence of excess emission surrounding the image of the star.

The second unusual feature of the sight line to NGC~3783 is that it penetrates an outer edge of 
the Antlia supernova remnant (G275.5+18.4) (McCullough et al. 2002 ; Fesen et al. 2021).  As 
we mentioned briefly in Section~\ref{sec:intro}, shocked clouds within supernova remnants are 
known to exhibit elevated pressures.  We examined a data cube of the results from the {\it 
Wisconsin H-Alpha Mapper\/} (WHAM) (Haffner et al. 2010) in an attempt to identify the radial 
velocity of the H$\alpha$ emission in the general location of NGC~3783, but we found that the 
shell was visible over a range $-35 < v_\odot < +35\kms$, which precludes our being able to 
make a unique identification among our three chosen \ion{C}{1} velocity intervals for the 
material within the SNR interior.

\subsection{A General Comparison with the Results of JT11}\label{sec:general_comparison}

\begin{figure}[b]
\gridline{\fig{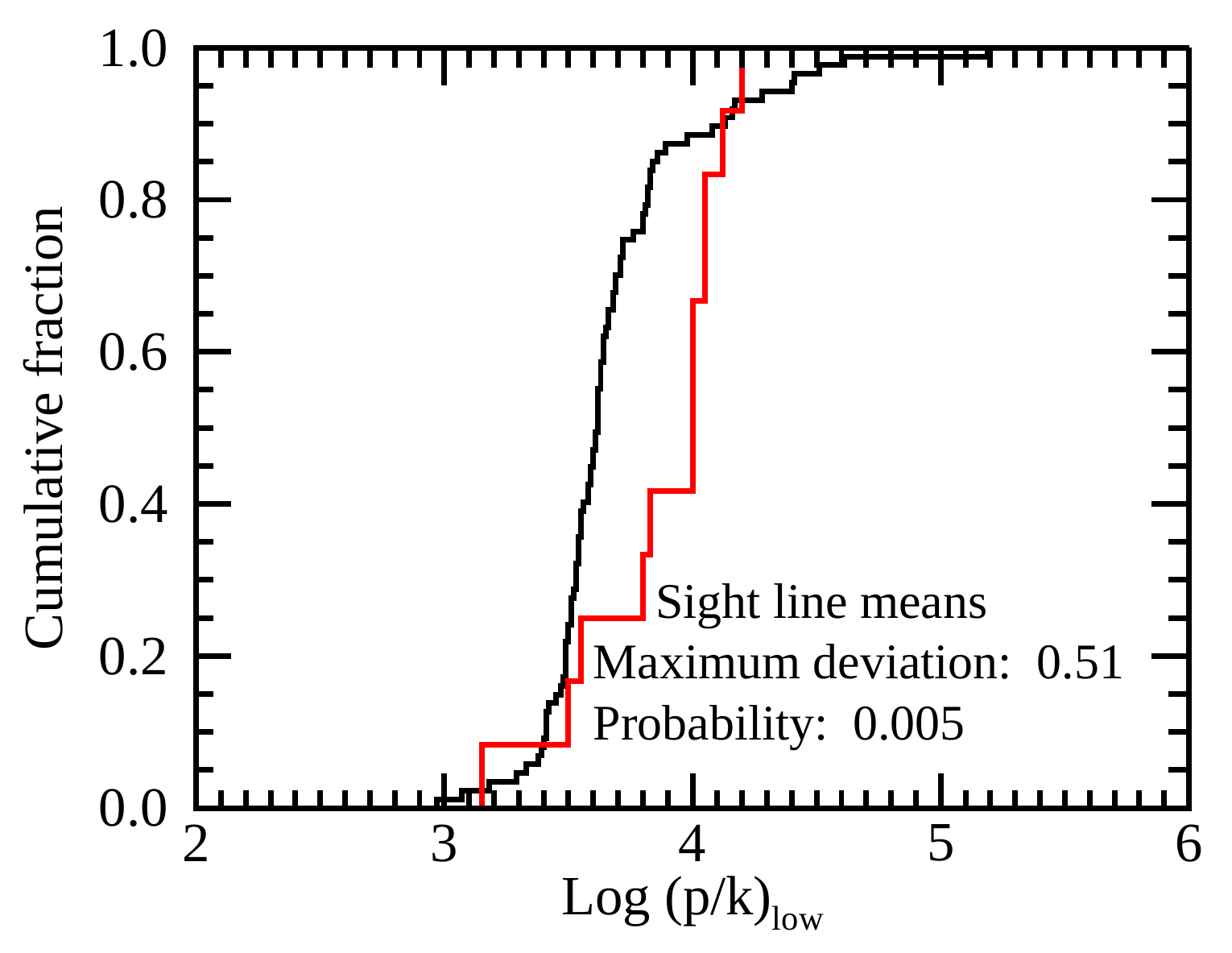}{0.45\textwidth}{(a)}
               \fig{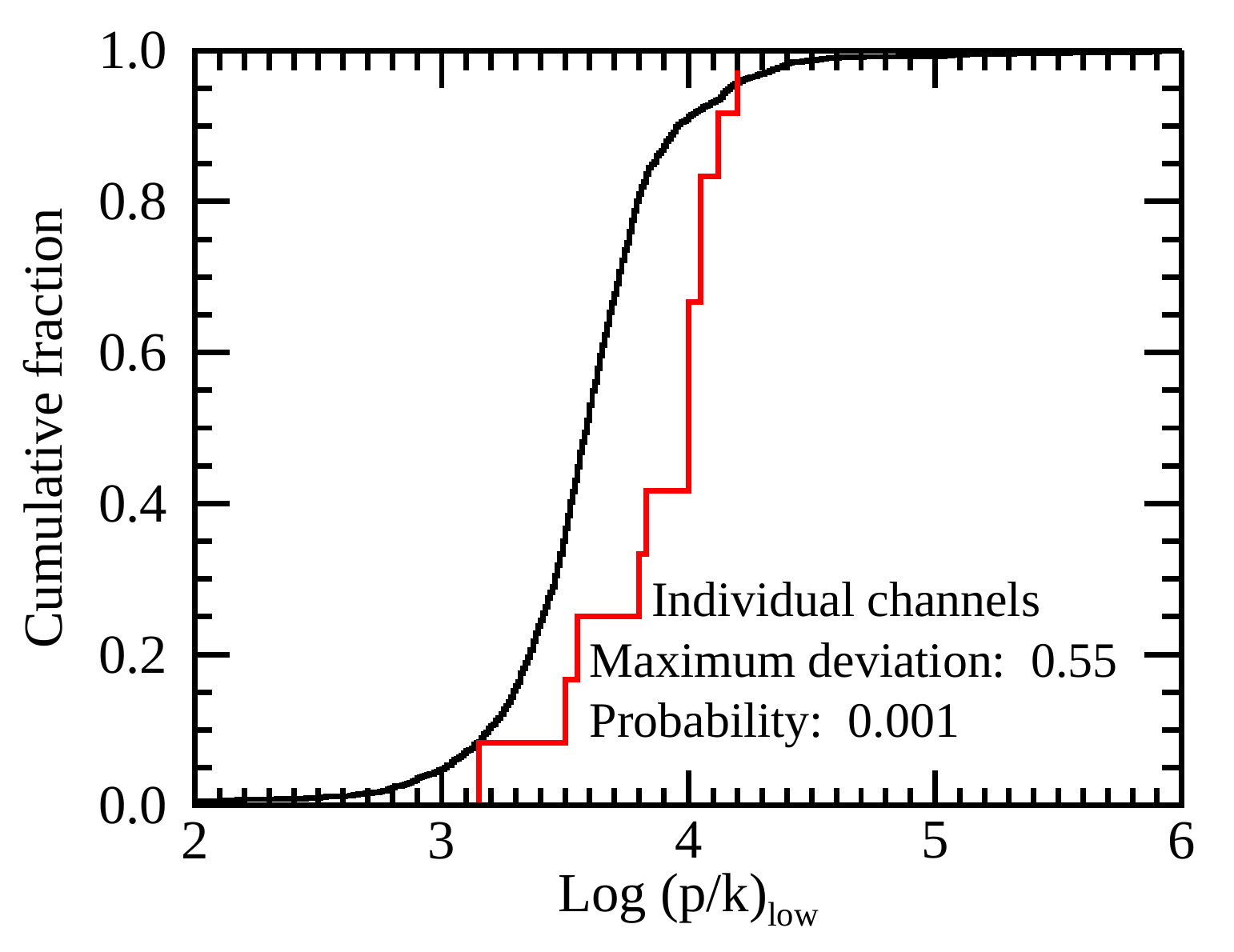}{0.45\textwidth}{(b)}}
\gridline{\fig{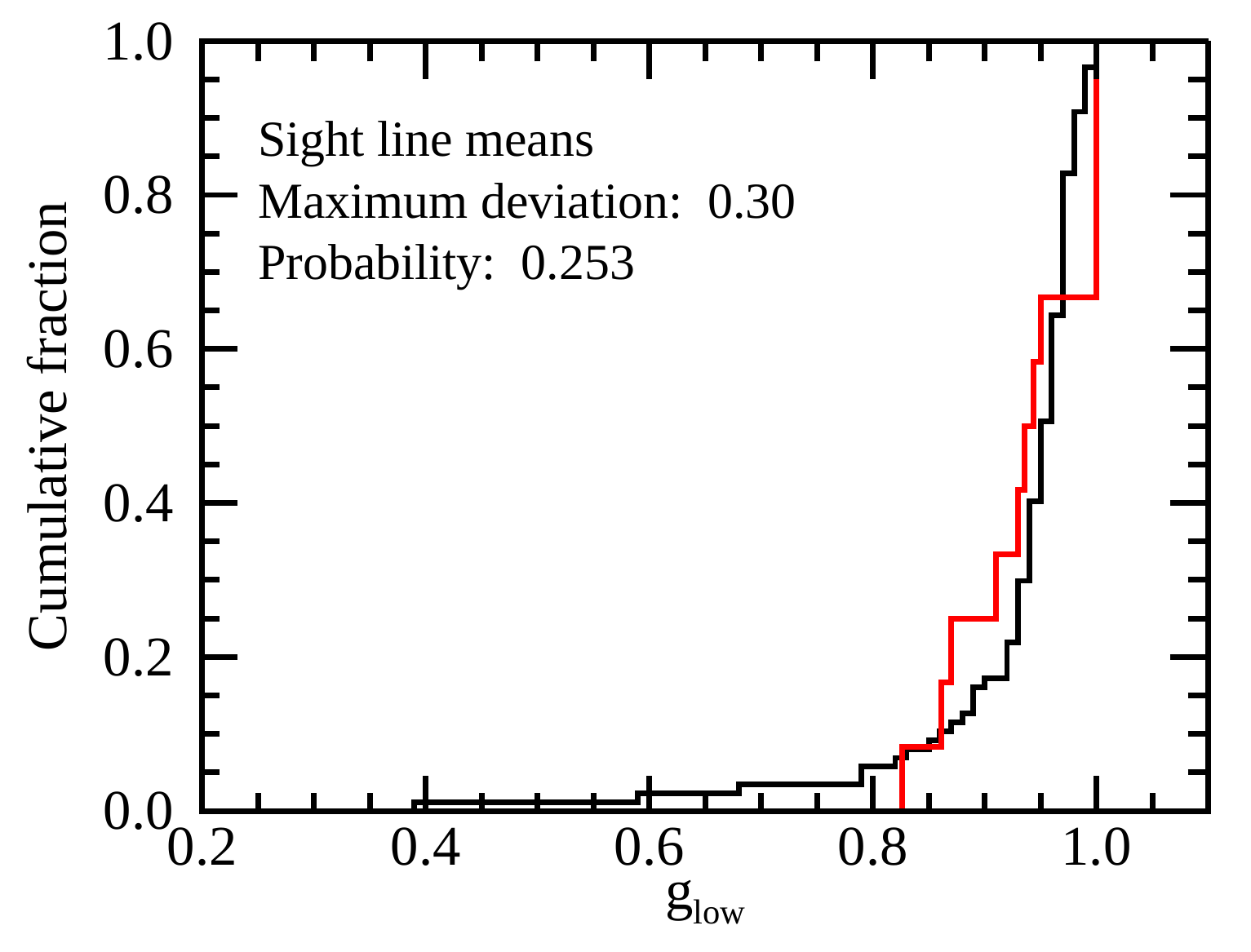}{0.45\textwidth}{(c)}
               \fig{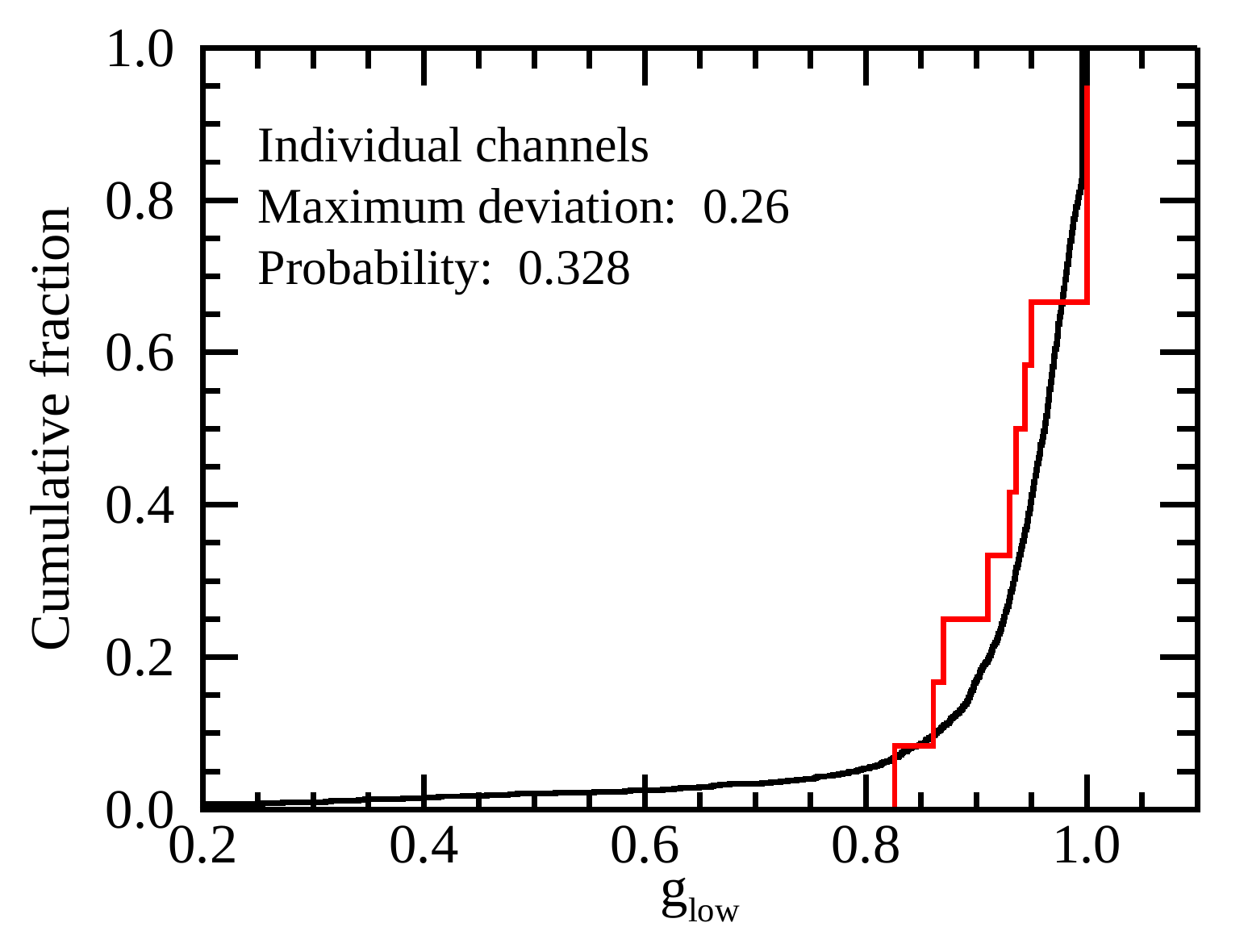}{0.45\textwidth}{(d)}}
\caption{Comparisons of the cumulative distribution functions for our sample (red) and those 
of JT11 (black) for $\log (p/k)_{\rm low}$ belonging to the dominant, low pressure fractions of 
\ion{C}{1} (panels (a) and (b)), and the relative amounts $g_{\rm low}$ of \ion{C}{1} in this 
component (panels (c) and (d)).  Panels (a) and (c) show comparisons with the mean values for 
the different sight lines listed in Table~3 of JT11, while panels (b) and (d) compare our results 
against the multitude of narrow velocity channels shown in Table~4 in that study.  In each panel 
we specify two quantities: (1) the maximum deviation in the vertical direction between the two 
distributions and (2) the resulting K-S test probabilities that the two outcomes are drawn from a 
single distribution.\label{fig:cumulatives}}
\end{figure}

For the reasons that we have just stated, we have elected to exclude the sight line toward 
NGC~3783.  Of the remaining results, for each case we can interpret a representative $\log 
(p/k)_{\rm low}$ of the majority of the \ion{C}{1} atoms, along with their fraction of the total 
$g_{\rm low}$ at these low pressures.\footnote{Following a procedure similar to that adopted 
by JT11, we imagine the projection along a line that starts from an assumed locus for the high 
pressure \ion{C}{1} atoms situated at $(f_1,f_2)=(0.38,0.49)$, passes through the location of 
the highest probability density outcome (depicted as red regions in 
Figures~\ref{fig:prob_plots1} and \ref{fig:prob_plots2}) and then intercepts the low pressure 
curve at a certain value of $\log (p/k)$ for an assumed CNM temperature $T\sim 100\,$K.  The 
fraction of \ion{C}{1} at high pressure $1-g_{\rm low}$ is proportional to the displacement 
along this line away from the curve.  It is important to understand that this quantity is not 
equivalent to the mass fraction of the gas at high pressures because there are no corrections 
for shifts in the ionization equilibrium of carbon.}  Table~\ref{tbl:outcomes} summarizes our 
results for these two quantities.  We identify the remaining $1-g_{\rm low}$ amount of gas in 
terms of high pressure contributions that are less well defined but still considerably above the 
low pressure range ($p/k\lt 20,000\cmk$).

\begin{deluxetable}{
r	
c	
c	
}
\tablewidth{0pt}
\tablecolumns{3}
\tablecaption{Outcomes from the Level Populations\label{tbl:outcomes}}
\tablehead{
\colhead{Target} & \colhead{$\log (p/k)_{\rm low}$} & \colhead{$g_{\rm 
low}$\tablenotemark{a}}
}
\startdata
\cutinhead{Distant Objects}
\object{3C273}&3.5&0.910\\
\object{HS 0624+6907}&3.8\tablenotemark{b}&1.00\tablenotemark{b}\\
&3.15\tablenotemark{c}&0.944\tablenotemark{c}\\
\cutinhead{Magellanic Cloud Stars}
\object{Sk $-$67 191}&4.0&1.00\\
\object{Sk $-$68 15}&4.2&0.930\\
\object{Sk $-$69 104}&4.12&1.00\\
\object{Sk $-$71 45}&4.05&1.00\\
\object{AzV 266}&4.05&0.861\\
\object{HD 5980}&3.55&0.936\\
\object{AzV 104}&4.00&0.826\\
\object{HD 5045}&4.00&0.870\\
\object{AzV 70}&3.83&0.950\\
\enddata
\tablenotetext{a}{Proportion of \ion{C}{1} at a pressure $(p/k)_{\rm low}$.}
\tablenotetext{b}{Component~1 (Velocity range: $-11.25$ to $-0.75$\kms).}
\tablenotetext{c}{Component~2 (Velocity range: $-0.75$ to 14.25\kms).}
\end{deluxetable}

Our main goal is to ascertain if there is any difference between the results obtained in this 
study and those derived by JT11.  An effective way to compare the two and assess whether or 
not they differ in a statistically significant manner is to apply a Kolmogorov-Smirnov (K-S) test 
on relevant combinations of two cumulative distribution functions.  Before doing so, we must 
recognize that there are differences between the way we measured pressures and the methods 
employed by JT11.  JT11 presented the measurements of $\log (p/k)_{\rm low}$ and $g_{\rm 
low}$ in two forms: one was a weighted mean of the quantities of the two variables over all 
velocities for a given line of sight, as listed in their Table~3, while the other was a set of explicit 
outcomes in individual velocity channels having widths of only $0.5\kms$ (Table~4).  Our 
analysis may be regarded as being intermediate between these approaches: for some sight 
lines we isolated velocity intervals where significant changes seemed to occur and analyzed 
them separately, but we always evaluated weighted means of the results over moderately wide 
velocity intervals to increase the accuracy of the outcomes, as we show in Column 5 of 
Table~\ref{tbl:targets}.  Since our approach does not exactly match the ones of JT11, we will 
compare our outcomes to both of those defined by JT11.

The results of the K-S tests are shown in Fig.~\ref{fig:cumulatives}.  We conclude that the 
cumulative distribution of the results for $\log (p/k)_{\rm low}$ in our study is significantly 
different from both of the two measurement outcomes of JT11.  The behavior of the 
distributions indicates a prevalence of somewhat higher pressures in our survey.  For the 
relative proportions of high and low pressure gas, represented by distributions of $g_{\rm 
low}$, the differences in the distributions are not significant.

If we examine the cases displayed in Fig.~\ref{fig:prob_plots2}, there is a suggestion that 
$g_{\rm low}$ is close to 100\% for the Galactic gas in front of the LMC (panels a-d in the 
figure), while for the SMC this quantity averages slightly less than 90\% (remaining panels).  We 
cannot identify any reason why this should be so, other than the existence of natural changes 
from one region of the sky to the next, and moreover the number of cases may be too small to 
claim that the effect is real.

\section{Discussion and Conclusions}\label{sec:discussion}

After considering the preliminary remarks that we stated in Section~\ref{sec:origin}, it is 
evident that our comparison of sight lines away from stars and those toward them yielded an 
unexpected result.  Before conducting the present survey, we had anticipated that the 
pressures found by JT11 toward early-type stars might be somewhat higher than those that 
were along sight lines that avoided such stars.  Instead, we found the opposite to be true.  
Hence the present survey indicates that the distribution of $\log (p/k)_{\rm low}$ found in the 
study of JT11 was not appreciably influenced by various possible forms of pressure 
enhancement near the stars.  This conclusion does not apply to the relatively small number of 
cases where the starlight density is clearly far above the average value with $g_{\rm 
low}\lesssim 0.5$ (i.e., the red points shown in Fig.~\ref{fig:JT11_plot}).

A different issue is our comparison of $g_{\rm low}$ for the two surveys.  While the significance 
levels for the two distributions being different from each other are quite poor from the 
perspective of a K-S test, the figures appear to indicate that over a range $0.85\lt g_{\rm low}\lt 
0.95$ the cumulative distribution for the current results shows a marginally faster rise, which 
indicates that about 60\% of our cases show a slightly larger contribution from very high 
pressure gas.  The proposition that the high pressure contributions are about the same as or 
perhaps slightly higher than what JT11 found reinforces the findings from 21-cm absorption 
studies that there are small regions (TSAS) with extraordinarily high densities and pressures in 
the general interstellar medium.

We must express an important caveat about the meaning of the significance levels of the K-S 
tests and the apparent differences in the cumulative distributions depicted in 
Fig.~\ref{fig:cumulatives}.  Our sample does not represent truly random directions in the sky.  
The Magellanic Cloud observations are tightly grouped, with many separations of order or less 
than one degree within in each cloud.  If the coherence lengths for $\log(p/k)$ or $g_{\rm 
low}$ are predominantly larger than these separations, the observations should not be 
independent of each other.  This condition will weaken the significance levels.  Unfortunately, 
we were unable to obtain a more diverse sample of sightlines by identifying in the public 
archive additional AGN or QSO spectra that showed measurable \ion{C}{1} features recorded by 
the STIS echelle spectrograph.  While observations by the Cosmic Origins Spectrograph (COS) on 
HST exist for many extragalactic targets, the lower resolution and broad wings of the COS line 
spread function preclude our obtaining reliable results for the \ion{C}{1} absorptions. 

Finally, we reiterate a point that we made in Section~\ref{sec:extragalactic}, which is that the 
sight lines in the current survey sample less gas than those generally represented by the 
measurements of JT11.  This difference may explain why we see slightly higher pressures.  For 
both surveys, the sight lines penetrate the outer boundary of the Local Bubble, an irregular void 
in the cool gas that is approximately centered on our location and which has a radius of about 
100~pc (Lallement et al. 2013 ; Liu et al. 2017).  This bubble is believed to have been created by 
multiple supernova explosions over a period of about $2-15$~Myr ago (Fuchs et al. 2006 ; 
Schulreich et al. 2018).  Its boundary may consist of an arrangement of clouds or shells holding 
cool gas at somewhat elevated thermal pressures (Kim et al. 2017).  Because the present survey 
samples less gas overall, the material at the Local Bubble boundary could conceivably have a 
greater influence on our results compared to the samples in the survey conducted by JT11.

\acknowledgments
We thank Eve Ostriker for her useful insights on some of the topics that we considered.
This research was based on observations with the NASA/ESA Hubble Space Telescope obtained 
from the {\it Mikulski Archive for Space Telescopes\/} (MAST) maintained at the Space 
Telescope Science Institute (STScI), which is operated by the Association of Universities for 
Research in Astronomy, Incorporated, under NASA contract NAS5-26555.  This study was 
partially supported by grants HST-GO-15321.002-A to Princeton University and
HST-GO-15321.001-A to the University of Massachusetts from the STScI under NASA contract 
NAS5-26555. This research has made use of the NASA/IPAC Infrared Science Archive, which is 
funded by the National Aeronautics and Space Administration and operated by the California 
Institute of Technology.  The Wisconsin H$\alpha$ Mapper and its H$\alpha$ Sky Survey have 
been funded primarily by the National Science Foundation. The facility was designed and built 
with the help of the University of Wisconsin Graduate School, Physical Sciences Lab, and Space 
Astronomy Lab. NOAO staff at Kitt Peak and Cerro Tololo provided on-site support for its 
remote operation.

\facilities{HST (STIS), IRSA, Spitzer, WHAM}

\end{document}